\documentclass[%
superscriptaddress,
amsmath,amssymb,
 aps,
twocolumn,
]{revtex4}
\usepackage{amssymb}

\renewcommand{\thesubsection}{\thesection.\arabic{subsection}}

\usepackage{enumitem}
\usepackage{graphicx}
\usepackage{dcolumn}
\usepackage{bm}
\makeatother
\graphicspath{ {figs_New/} }
\usepackage{enumitem}

\newcommand{\R}{\mathbb{R}}
\newcommand{\grad}{\rm grad}
\usepackage{color}

\begin{document}


\title{Curvilinear polyhedra  as dynamical arenas, illustrated by  \\
	an example of self-organized locomotion}

\author{Shankar Ghosh$ ^1 $, A. P. Merin$ ^1 $, S. Bhattacharya}





 \affiliation{Department of Condensed Matter Physics and Materials Science}

\author{Nitin Nitsure}%

 \affiliation{School of Mathematics \\ Tata Institute of Fundamental Research, Mumbai 400005, India}




\begin{abstract} 
Experiment shows that 
dumbbells, placed inside a tilted hollow cylindrical drum that rotates slowly around its axis, climb uphill by forming dynamically stable pairs, seemingly against the pull of gravity.
Analysis of this experiment shows that the dynamics takes place in an underlying space which is a curvilinear polyhedron inside a six dimensional manifold, carved out by unilateral constraints that arise from the 
non-interpenetrability of the dumbbells.
The  energetics over this polyhedron localizes the configuration point within the close proximity of a corner of 
the polyhedron. This results into a strong entrapment, which provides the configuration of the dumbbells with its observed shape that leads to its functionality -- uphill locomotion. The stability of the configuration
is a consequence of the strong entrapment in the corner of the polyhedron. 
\end{abstract}






\maketitle

\section{Introduction}
Unilateral constraints, that is, constraints in the form of 
inequalities $f_i(q_j, \dot{q_j}) \ge 0$
in generalized coordinates and velocities $(q_j, \dot{q_j})$, are common in everyday life. They often
arise from the non-interpenetrability of physical objects, which is 
the context of contact mechanics.
Such constraints are inconvenient in the framework of mechanics on smooth 
manifolds.   
When the unilateral constraints involve only $ q_i $'s, as in contact mechanics,
they can be accounted for as the effects of additional ad hoc sharply 
rising potentials, that is, in terms of mechanical deformations at the points of
contact \cite{meirovitch2010methods,pfeiffer_multibody_2004, johnson1987contact}. 
In this paper, we instead look at the curvilinear polyhedra
in suitable generalized coordinates that are carved out by the unilateral constraints. 
The behaviour of the mechanical systems plays out on these polyhedra. We find that 
essential qualitative aspects of the behaviour, such as stable entrapment and bifurcation are closely related to the local geometry 
near the corners of the polyhedra. 

As an illustration of this paradigm, we study an experimental example. When a number of identical dumbbells are  placed inside a tilted hollow cylindrical drum that rotates slowly around its axis, they climb upwards by forming dynamically stable pairs (dyads), seemingly against the pull of gravity. It is a surprise when objects move in a direction opposite to the apparent force applied to them. It is even more surprising when such behaviour is displayed 
not by single objects, but only by pairs of objects which behave as a unit that is dynamically stable  without any mutual attraction between the constituents. In this paper, we first 
identify a polyhedron in a certain manifold that naturally arises because of 
non-interpenetrability of the dumbbells, and then we show how energetics over the corners of this 
polyhedron can explain the observed behaviour of the dumbbells.

This paper is arranged as follows. The experimental arrangement is described in section 2,
and the observations are described in section 3. The section 4 explains the behaviour of a single dumbbell.
The physical description of the curvilinear polyhedron and its faces, which underlie the theory for dubbell pairs, 
is given in section 5. The section 6 treats energetics over this polyhedron, and identifies its
local energy minimizing locus. With this preparation, the section 7 completes the explanation of the observed facts about 
dumbbell pairs. This is followed by general conclusions and speculations in section 8. The 
appendix \ref{appn_friction} gives a quantitative analysis of the frictional response of dumbbells.
The appendices \ref{appn_polyhedron} and \ref{appn_enrgy} 
contain mathematical details used in the paper.

\begin{figure}[t]
	\centering
	\includegraphics[width=0.9\linewidth]{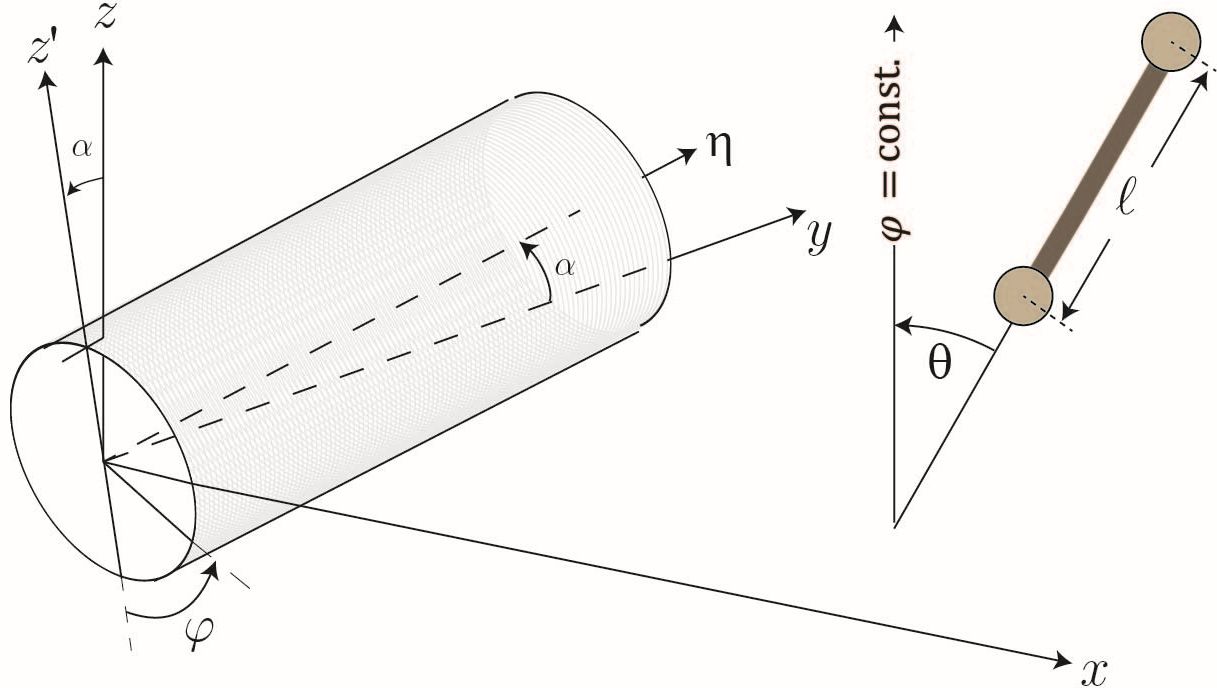}
	\caption{A glass cylinder that is tiled at an angle $\alpha$ with respect to the horizontal is made to rotate about its axis with a constant angular speed $\omega$. The inset shows an amplified dumbbell of length $\ell$ that is tilted at an angle $\theta$ from the meridian $\varphi=constant$. The angle $\theta $ (the heading) is positive for the dumbbell that is depicted. 
	}
	\label{fig:Expt_schematic}
\end{figure}

\section{The experimental arrangement} \label{sec:Geometry}

\begin{figure}[tb]
	\centering
	\includegraphics[width=1\linewidth]{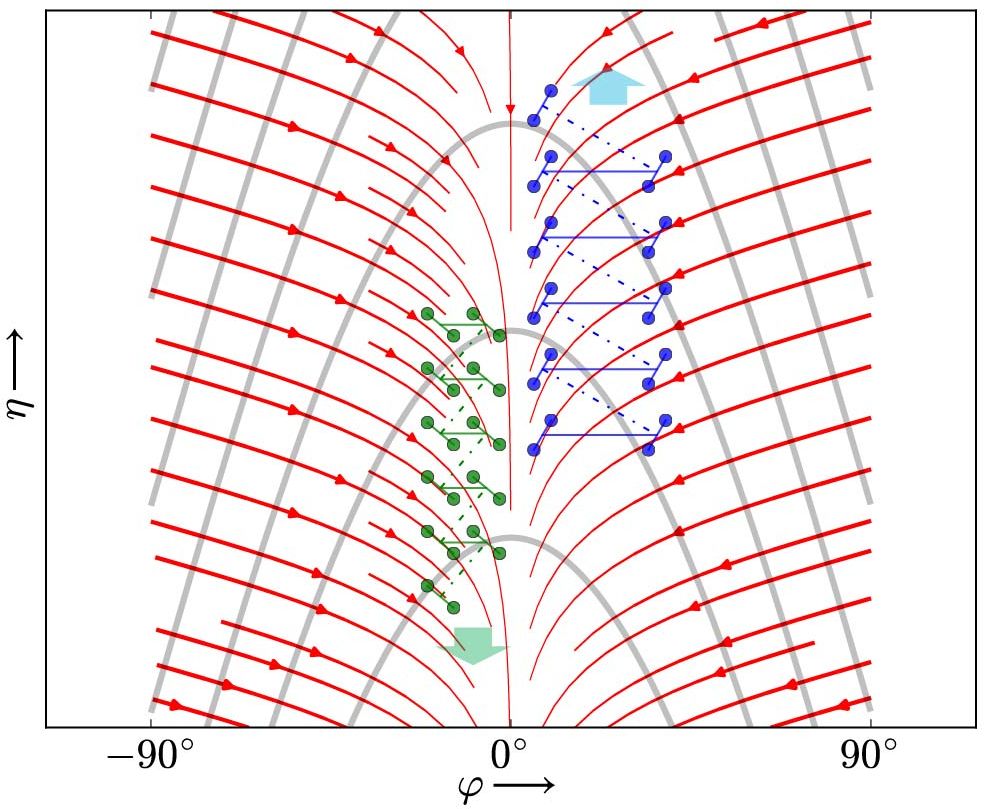}
	\caption{The rectangle represents the the region $\Omega$ on the surface of the cylinder, in terms of its intrinsic coordinates $\varphi$ and $\eta$. The level curves of constant height $z$ are shown in grey and the flow lines of $-{\rm grad}(z)$ are shown in red. The blue and the green zigzag trajectories are representative  paths followed by isolated dumbbells which have a constant positive or negative heading, respectively.
	The blue zigzag climbs upwards in $\eta$ while the green zigzag climbs downwards in $\eta$.}
	\label{fig:coordinates}
\end{figure}

  A cylindrical drum of radius $R\approx 125\,\mbox{mm}$ made of glass,   that is tiled at an angle $\alpha$ ( $ \approx 7^{\circ}$)  with respect to the horizontal,  is made to rotate about its axis at a constant angular speed $\omega$ ($\approx 0.01$ radians/sec). The dumbbells used in the experiment are made of two identical spherical balls rigidly joined by a  cylindrical rod in a symmetric manner.
   The distance between the centers of the balls will be denoted by $\ell$, which we will call as the \textit{length} of the dumbbell. The radius of the balls, which is $ \sim 3 \,\mbox{mm} $, satisfies $r \le \ell/2$. Moreover, the radius $t$ of the rod is significantly smaller than the radius  of the balls. 
 If $r = \ell/2$, the dumbbell will appear as a pair of spheres glued together. The dumbbells used in the experiments are made of plastic.  Their  size  is tiny compared to the size of the cylinder ($\ell/R \le 1/10$).

The intrinsic coordinates $\varphi$ (`azimuthal angle')  and $\eta$ (`cylindrical altitude') on the cylinder (see Fig. \ref{fig:Expt_schematic}) are defined in terms of 
the laboratory coordinates $x,y,z$ (with $z$ the vertical coordinate)  
by the equations 
\begin{eqnarray*}
\varphi & = &-\tan^{-1}\left({x\over -y\sin\alpha+ z \cos\alpha}
	\right) \mbox{ and } \\
	\eta & = & y\cos\alpha + z\sin\alpha.
	\end{eqnarray*}
The bottom of the cylinder is at $\eta=0$.
Infinitesimal distance on the surface of the cylinder is given in terms of these coordinates
by $ds^2 = R^2 d\varphi^2 + d\eta^2$. The
geodesics on the surface of the cylinder are given by linear equations $a\varphi + b\eta +c = 0$. These are helices in general and as special cases they include the straight lines $\varphi = \mbox{ constant}$  (referred to as the meridians) and the circles $\eta =\mbox{ constant}$ (referred to as the latitudes). Note that $\varphi=0$ is the lowest straight line on the curved 
surface of the cylinder.

Unlike latitude and longitude coordinates on the earth, the coordinates $\varphi,\eta$ are not to be
regarded as rotating with the cylinder. Consequently, 
the rotation of the cylinder carries a 
point $(\varphi, \eta)$ to the new point $(\varphi + \omega t, \eta)$  after a time $t$, by moving along a latitude.

The physical height function $z$ on the cylinder is given in these coordinates by the formula 
 \begin{equation*}
  z=\eta \sin \alpha - R \cos \alpha \cos\varphi .
 \end{equation*}
Its level curves on the cylinder are depicted in black in Fig.\ref{fig:coordinates}.
The corresponding gradient vector field $F_T$ on the surface of the cylinder is given by 
\begin{equation}
F_T= -\grad(z) = -\cos\alpha\sin\varphi\, R^{-1} \frac{\partial}{\partial\varphi}  
-  \sin\alpha \,\frac{\partial}{\partial\eta}  
\label{Eqn:tangent_force}
\end{equation}
The flow lines of $F_T$ constitute the family of curves 
described by the differential equation 
$d\varphi /d\eta = (\sin\varphi)/(R\tan\alpha) $. 
These are depicted in red in Fig.\ref{fig:coordinates}.
The flow lines intersect the level curves orthogonally at all points.  The lowest 
meridian line $\varphi =0$ is one such flow line.

The region $\Omega$ on the cylinder, which is of relevance to the experiment, is the inner surface of the `lower half' of the cylinder, given in coordinate terms by $-\pi/2 < \varphi < \pi/2$ and $\eta > 0$.
The motion of the dumbbells in the experiment takes place in this region.
A dumbbell lying in $\Omega$ is described by coordinates $\varphi,\eta,\theta$.
Here, $\varphi,\eta$ give the location of the center of the dumbbell, and  
$\theta$ gives the angle from the 
 axial vector of the dumbbell to the meridian direction $\partial/\partial\eta$; see Fig.\ref{fig:Expt_schematic}. Because of the symmetry of the dumbbell, we identify
$\theta$ with $\theta + \pi$, so that $\theta$ becomes a periodic corrdinate  
with period $\pi$.
In this coordinate description of a dumbbell, we have quotiented out the 
angular coordinate which describes the rotation of a dumbbell around its own axis, as that 
is not used in what follows.
Suggested by a navigational analogy, we will call $\theta$ 
as the {\it heading} of the dumbbell.

\section{Experimental Observations}

\subsection{Isolated  dumbbells}

\begin{figure}[t]
	\centering
	\includegraphics[width=0.9\linewidth]{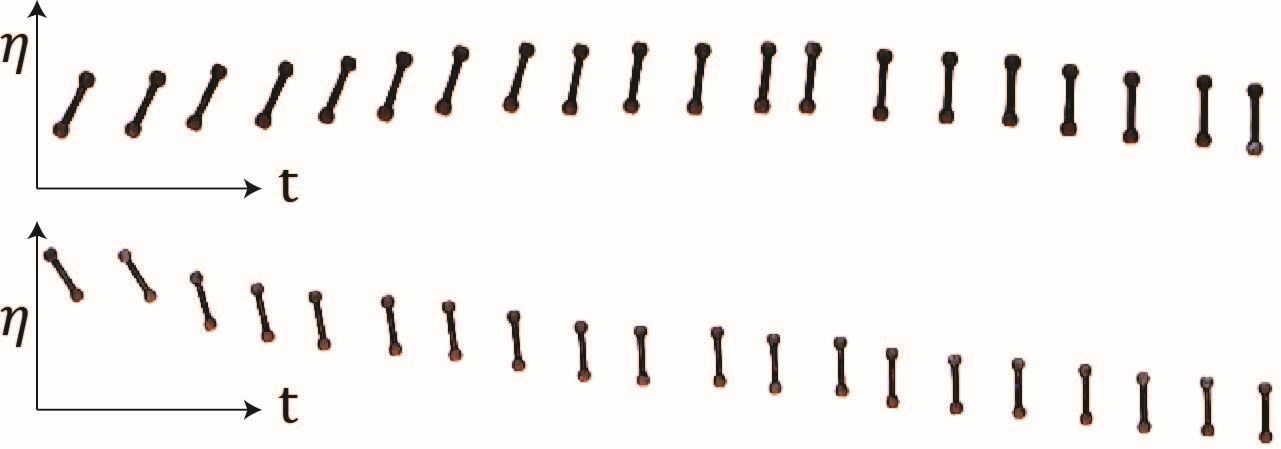} 
	\caption{Snapshots in time which capture the loss of  heading for dumbbells placed  in a tilted rotating cylinder ($ \alpha=7^{\circ} $) with an initial positive heading (top panel) and negative heading (bottom panel). The  physical parameters for the dumbbell are $\ell=14\,\mbox{mm},\,  r=3 \,\mbox{mm}, \, t=1.5 \,\mbox{mm}$.}
	\label{fig:single_dumbbell_1a}
\end{figure}

\begin{figure}[t]
	\centering
	\includegraphics[width=1\linewidth]{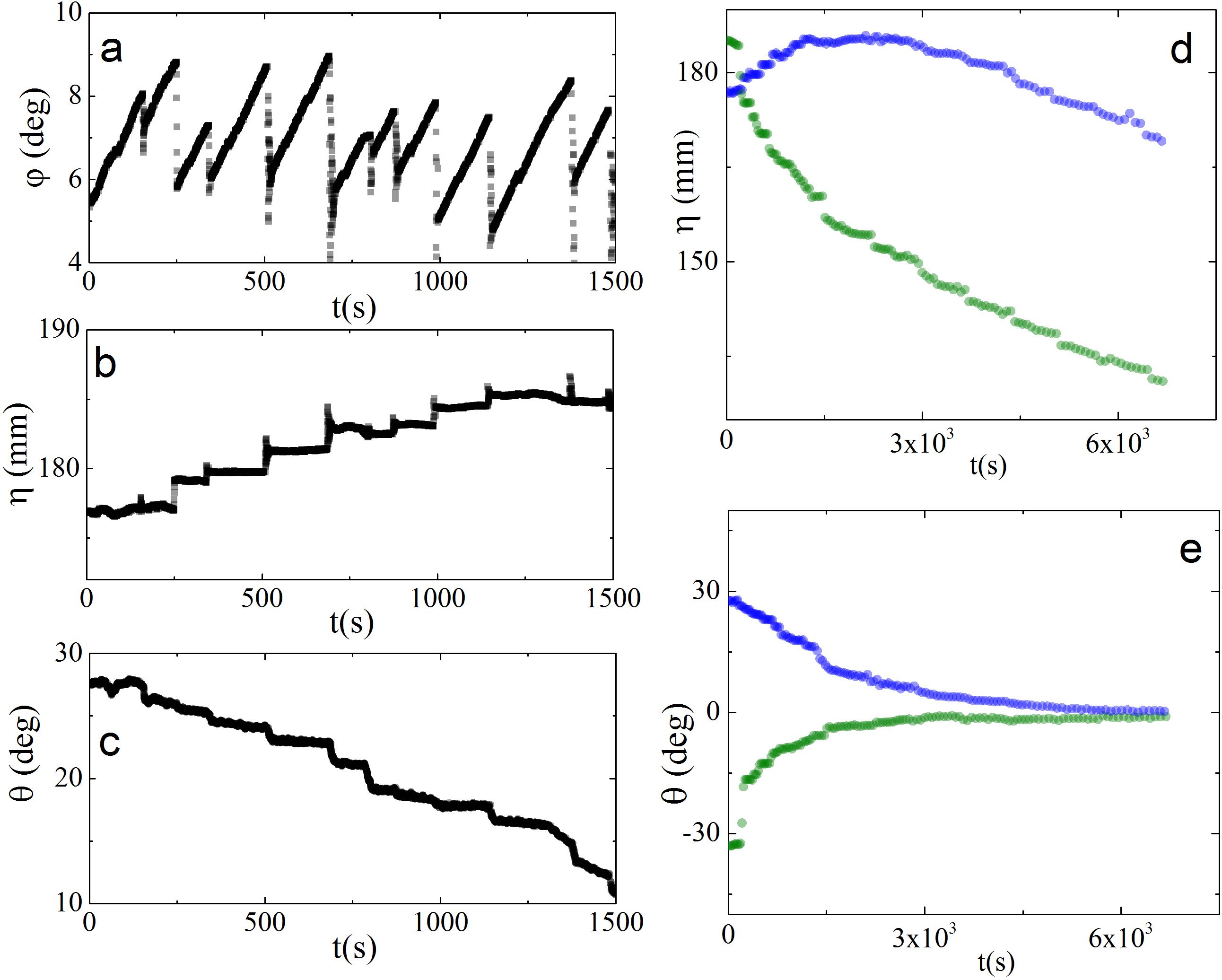} 
	\caption{The left half of this figure traces the history of a single dumbbell placed in a
		tilted rotating cylinder ($ \alpha=7^{\circ} $). The dumbbell had an initial positive heading. The 
		three panels (a), (b) and (c) on the left show how the values of $\varphi$, $\eta$ and $\theta$
		change with time.   The sharp drops in $\varphi$ seen in the panel (a) correspond to the rolling phase
		in the zigzag motion of the dumbbell. Cocomittently with these drops in $\varphi$, we observe sudden
		rises in $\eta$ as a result of rolling with a positive heading, as can be seen in the panel (b). Note that the drops in the value of the heading take place during the rolling phase, 
		as can be seen in the panel (c).  
		The right half of this figure traces the contrasting histories of two different isolated dumbbells: 
	the graphs in blue correspond to a dumbbell with an initial negative heading, and the 	 
		graphs in green correspond to a dumbbell with an initial positive heading. The panels
		(d) and (e)  on the right show how the values of $\eta$ and $\theta$
		change with time for these two dumbbells.   Note that for both the dumbbells, 
		$\theta$ goes to zero with time, and the value of $\eta$ begins to go down eventually.
		The  physical parameters for the dumbbell are $\ell=14\,\mbox{mm},\,  r=3 \,\mbox{mm}, \, t=1.5 \,\mbox{mm}$.}
	\label{fig:single_dumbbell_1}
\end{figure}

By imaging the motion of a single dumbbell placed in a tilted rotating cylinder, we observe the following.

\begin{enumerate}[leftmargin=*]
	 \item[$(i)$] A dumbbell with a small positive heading $0< \theta <  20^{\circ}$ when placed in the tilted rotating cylinder   moves up in $\eta$. Similarly a  dumbbell with a small negative heading $ - 20^{\circ}< \theta < 0$ when placed in the  cylinder   moves down  in $\eta$; see Fig.\ref{fig:single_dumbbell_1a}.

	  \item[$(ii)$] On closer inspection the trajectories for the cases described above are  seen to be zigzags  involving rolling and sticking phases (see  variations in time of $\varphi$ in Fig.\ref{fig:single_dumbbell_1}(a)  and  $\eta$ in Fig.\ref{fig:single_dumbbell_1}(b) along  such a trajectory). The speed at which a dumbbell rolls down in $\varphi$ 
	  is much greater than the speed
	  at which it gets carried up in $\varphi$ by the rotation of the cylinder: correspondingly,
	  the rolling phases appear as almost vertical segments in Fig.\ref{fig:single_dumbbell_1}(a).

	  \item[$(iii)$] The heading is not stable over time and tends to zero.  The changes are shown in Fig. \ref{fig:single_dumbbell_1a} and Fig.\ref{fig:single_dumbbell_1} (c).  The changing in heading takes place during the rolling arms of the zigzag (see  Fig.\ref{fig:single_dumbbell_1}). The width in $\varphi$ of these zigzags is about $5^{\circ}$. These zigzag  trajectories are  schematically shown (amplified for clarity) in Fig.\ref{fig:coordinates}.

	 \item[$(iv)$] A dumbbell with zero heading moves downwards in $\eta$ over time by intermittent slippages, keeping its heading nearly zero (see Fig. \ref{fig:single_dumbbell_1a} and  Fig.\ref{fig:single_dumbbell_1}(d) and (e)).
\end{enumerate}

From the above observations it follows that  a  dumbbell whose initial heading is between $-20^{\circ}$ to $20^{\circ}$ eventually moves to the bottom of the cylinder. If $|\theta|$ is larger, then the dumbbell rolls down
rapidly to the bottom of the cylinder.


The above observations continue to hold for $  \alpha \le 10^{\rm o}$. For a larger $\alpha$ as the increased downward force overcomes the force of sliding friction more easily, a dumbbell
rapidly descends to the bottom of the cylinder, regardless of its initial heading.

\subsection{Bunch of dumbbells } 

\begin{figure}[t]
	\centering
	\includegraphics[width=1\linewidth]{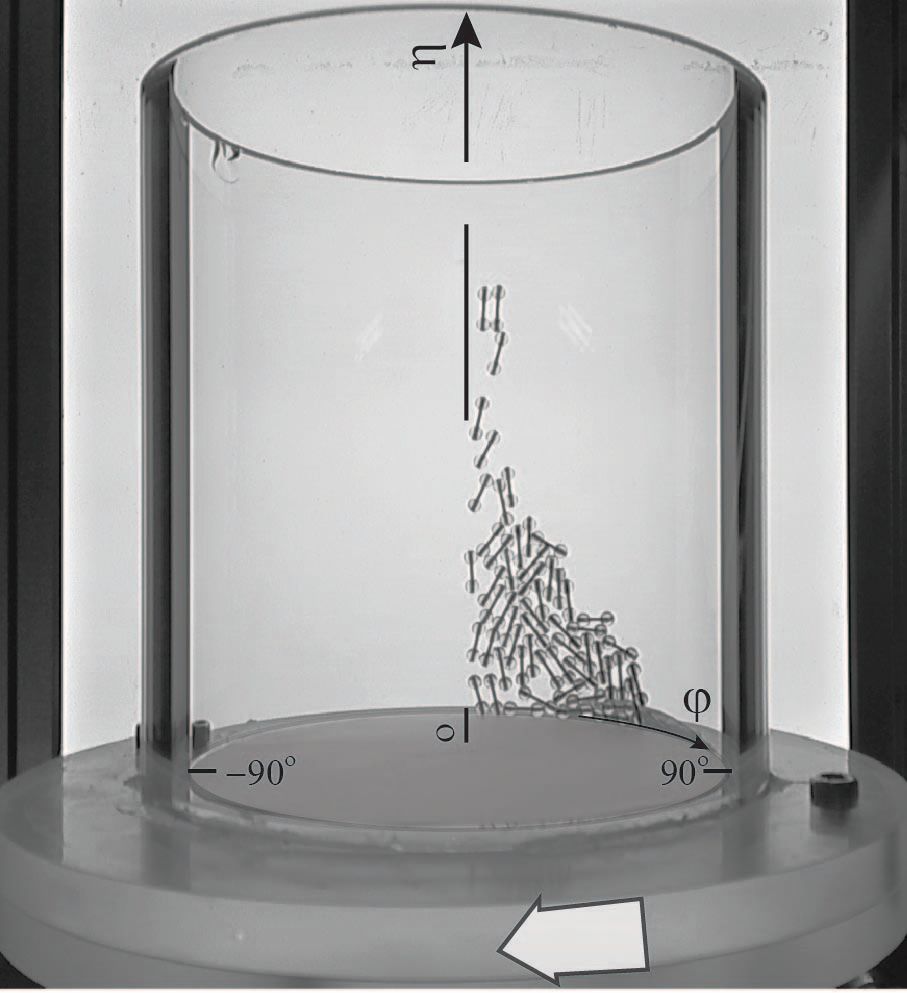}
	\caption{The image shows a partitioning of the dumbbells in a tilted  rotating cylinder ($ \alpha=7^{\circ} $)  into rollers and cluster of sliders. The rollers are found in the  region where $|\varphi|$ is small. The sliders form an interlocked structure located at higher values of $\varphi$.  The direction of rotation of the cylinder is marked by an arrow. The  physical parameters for the dumbbells are $\ell=8\,\mbox{mm},~  r=2\,\mbox{mm},~ t=0.5\,\mbox{mm}$.}
	\label{fig:bulk}
\end{figure}

\begin{figure}[t]
	\centering
	\includegraphics[width=1\linewidth]{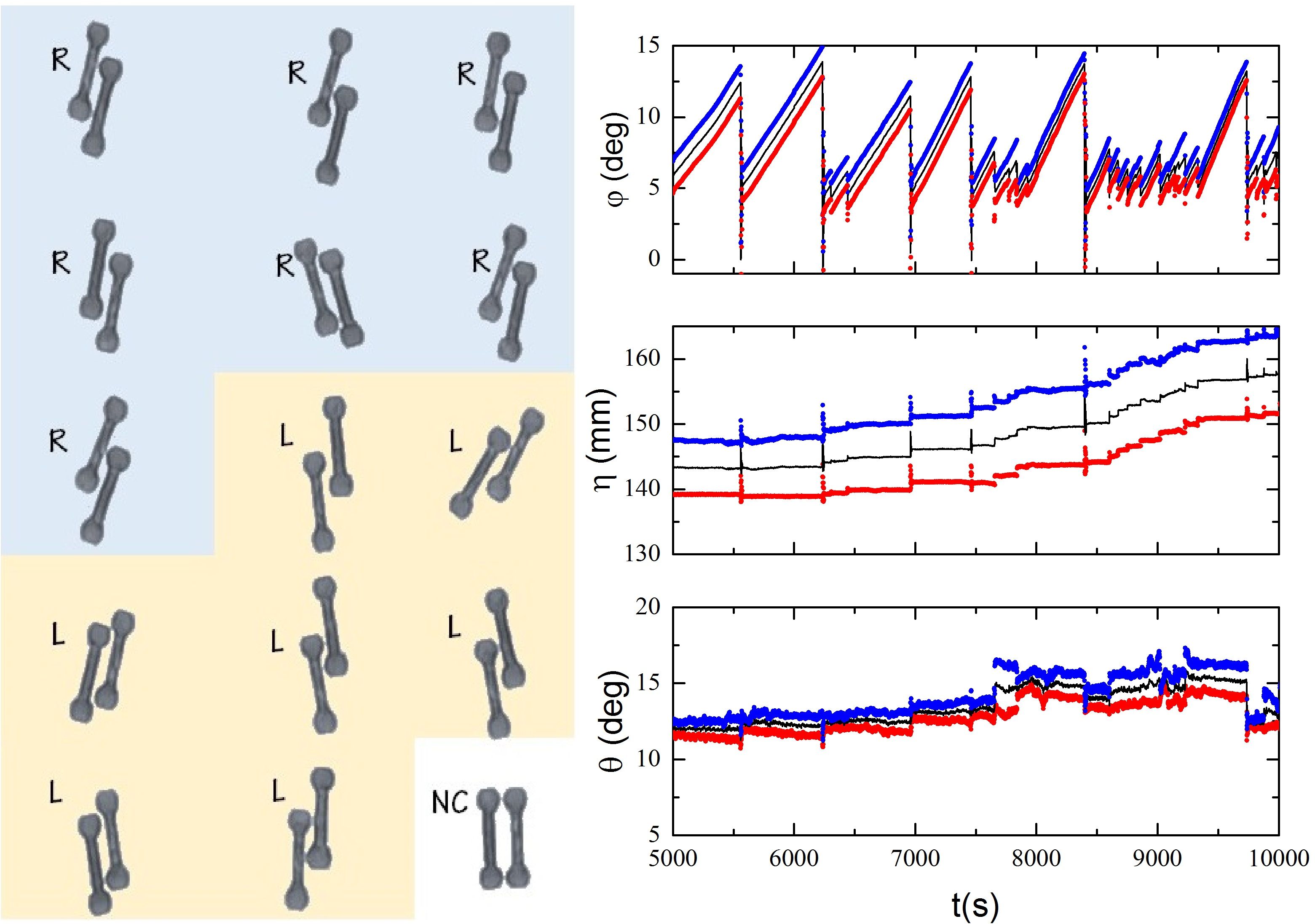}
	\caption{The left half of the figure show experimentally observed examples of left and right handed dyads, labelled L and R respectively.
	The pair labelled NC is a non-chiral unstable structure, of a kind that occasionally arises. 	
	 The right half of the figure traces the experimentally observed history of a right handed dyad in a tilted
	 rotating cylinder ($ \alpha=7^{\circ} $). The individual histories
	 of the two constituent dumbbells of this dyad are shown in blue and red respectively. It is
	 to be noted how the two dumbbells move in concert. The variation in time of the average values of the 
	 parameters is shown as a black curve. The  physical parameters for the dumbbell are $\ell=16\,\mbox{mm},~  r=3\,\mbox{mm}, ~ ~ t=1.5 \,\mbox{mm}$.	 }
	\label{fig:dyads_confg_data_comb}
\end{figure}

When a number of identical dumbbells are placed together as a bunch in the bottom of a tilted rotating cylinder, 
the bunch slides down to the lowest part of the tilted cylinder, where the flat bottom meets the 
curved surface. Subsequently, we observe the following.

%
%

\begin{figure}[t]
	\centering
	\includegraphics[width=1\linewidth]{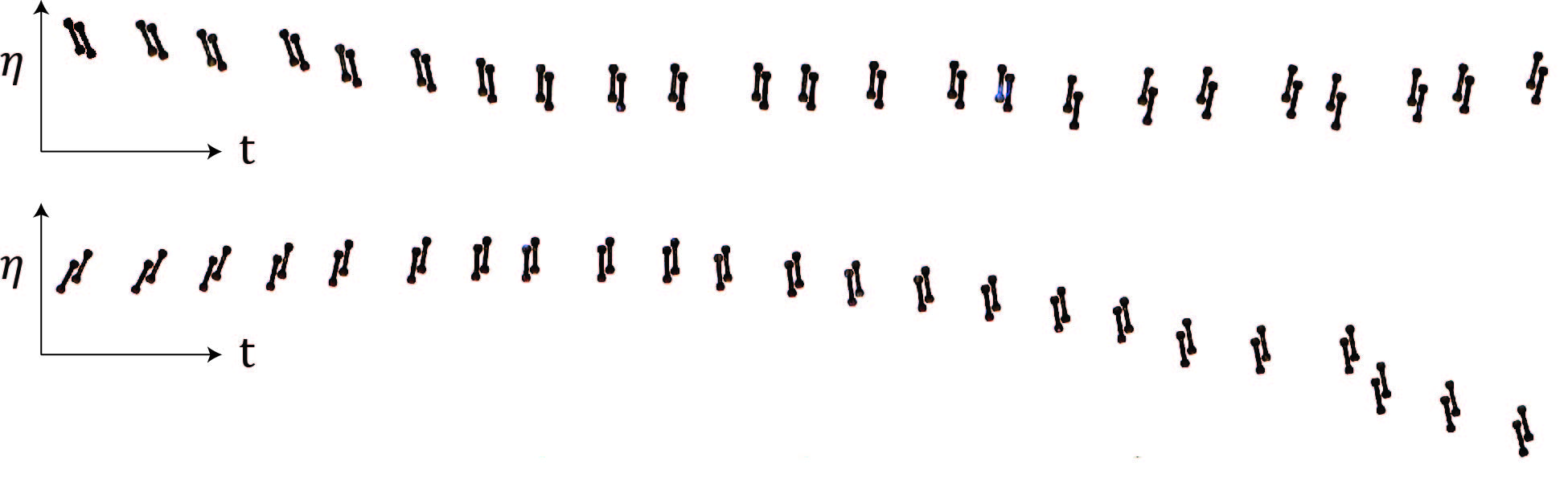}
	\caption{Snapshots in time which capture the stability  of  heading for dumbbell dyads  placed 
		in a tilted rotating cylinder ($ \alpha=7^{\circ} $). The top and bottom panels respectively show the observations for 
		right handed and a left handed dyad respectively. The observations deliberately begin which a right handed dyad with 
		an initial {\it negative} heading and left handed dyad with 
		an initial {\it positive} heading.
		It is to be noted how the headings of the right handed and left handed dumbbell dyads progressively
		become positive and negative respectively, and how the respective dyads eventually move in the $\eta$ or
		the $-\eta$ directions. 
		The  physical parameters for the dumbbells are $\ell=16\,\mbox{mm},~  r=3\,\mbox{mm},~ t=1.5\,\mbox{mm}$.}
	\label{fig:dyads_a}
\end{figure}

\begin{figure}[b]
	\centering
	\includegraphics[width=1\linewidth]{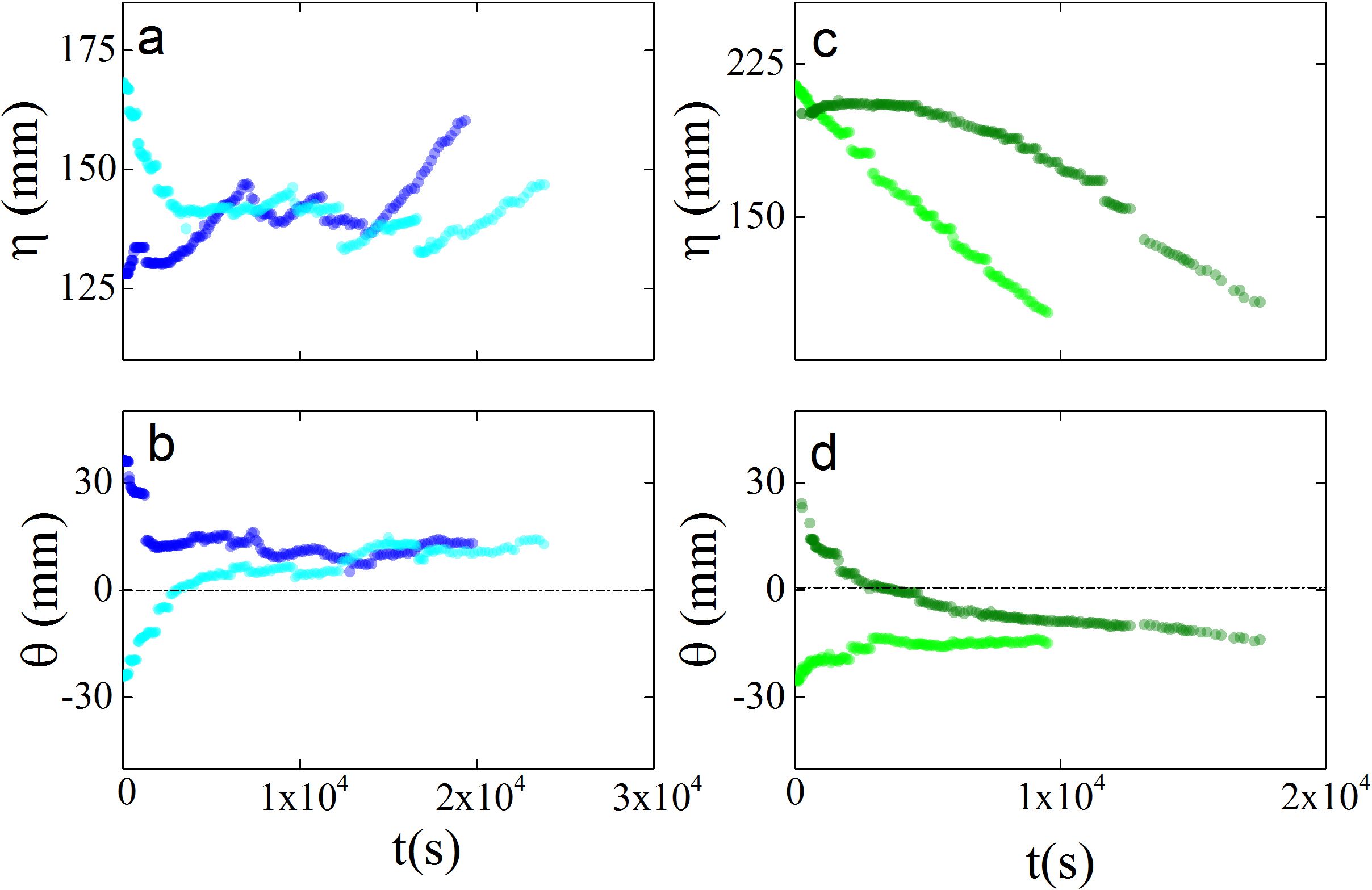}
	\caption{ Experimentally observed temporal variations of $\eta$ corresponding to  
the trajectories of two right handed dyads and two left handed dyad 
in a tilted  rotating cylinder ($ \alpha=7^{\circ} $) are shown in (a) and (c), respectively.  Their  corresponding variations in $\theta$ are plotted in (b) and (d) respectively. The data for any particular dyad is marked in a distinctive colour. For each chirality, one of the dyads has a positive initial heading and the other dyad has negative initial heading.
		 The  physical parameters for the dumbbells are
		 $\ell=16\,\mbox{mm},~  r=3\,\mbox{mm},~ t=1.5\,\mbox{mm}$.}
	\label{fig:dyads}
\end{figure}

\begin{figure}[b]
	\centering
	\includegraphics[width=1\linewidth]{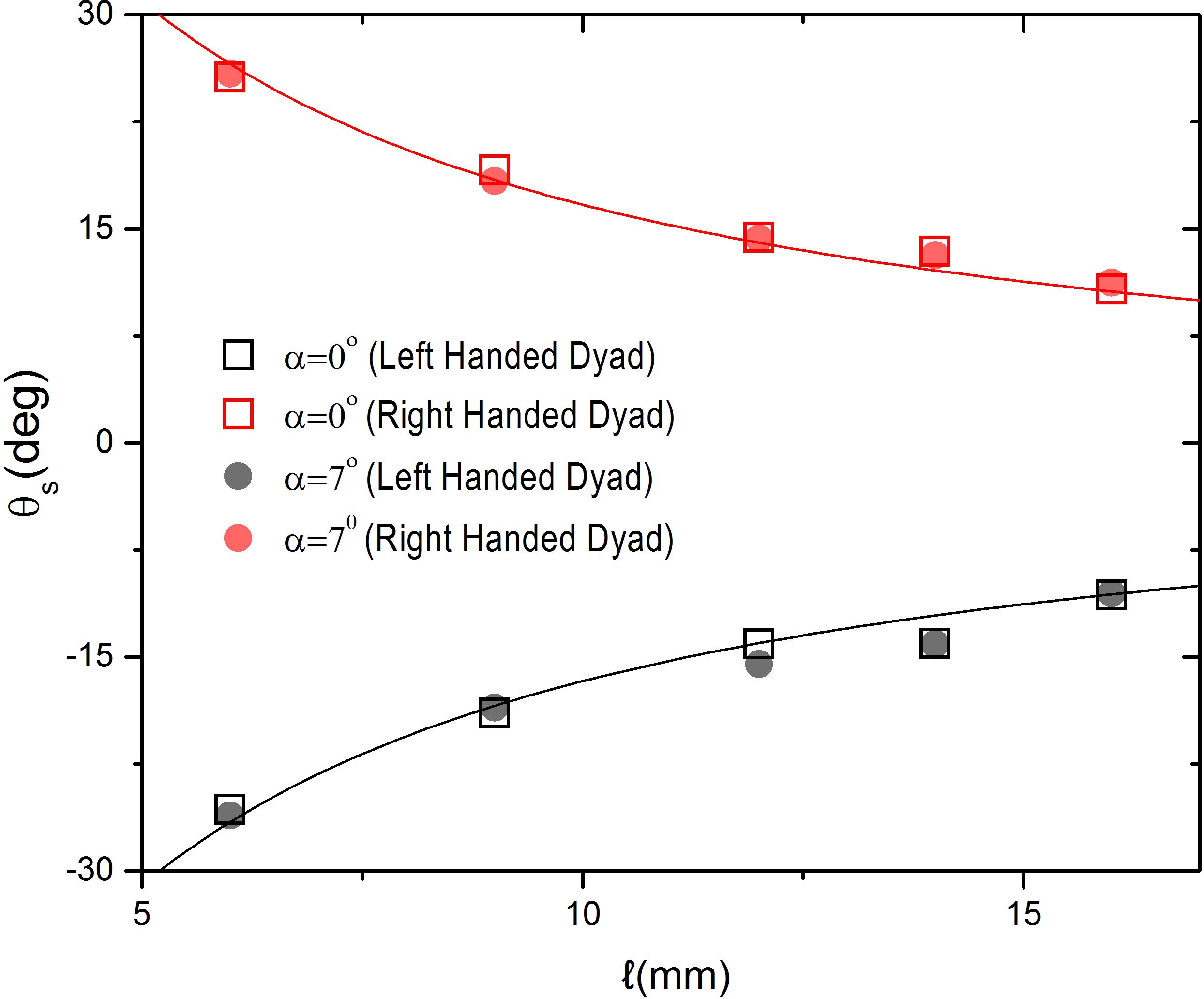}
	\caption{Experimentally observed variation of the stable heading $\theta_s$ as a function of $\ell$ for a right handed dyad are marked as red data points and for a left handed dyad as grey data points.  The open symbols are the data obtained for a horizontal cylinder ($ \alpha=0^{\circ} $) and the filled symbols correspond to the tilted cylinder  ($ \alpha=7^{\circ} $).	The red curve is the graph of $\theta_{s}=\arctan(r/\ell)$  and the black curve is the graph of $\theta_{s}=-\arctan(r/\ell)$ for a fixed value $r = 3\,\mbox{mm}$, which are the theoretical expected relations based on energy minimization derived in appendix \ref{appn_enrgy} for ideal dumbbells with weightless rods in a horizontal cylinder.}
	\label{fig:length_sat}
\end{figure}

 \begin{enumerate}[leftmargin=*]

\item[$(i)$] The bunch  gets  partitioned  into  a discrete set of rollers and a cluster of interlocked dumbbells that slide. The  rollers  occupy the region where  $|\varphi|$ is small with $\varphi$ mostly positive,  and the  interlocked sliding structures are located at a higher value of $\varphi$ (see Fig.\ref{fig:bulk} for a representative image of
how the dumbbell distribution begins to appear very soon after the start). 
This behavior is similar to that for rolling spheres which was described in Kumar. et. al.   \cite{kumar2015granular}.
There are occasional slippages and collisions, and headings are observed to get randomized. This process keeps producing isolated dumbbells with a small
positive heading.


\item[$(ii)$] Isolated rollers which have a small initial positive heading
begin to move up the cylinder, till eventually their headings become nearly zero and they begin to slip downwards. This is as described in section 3.1.

\item[$(iii)$] During the above process, descending
dumbbells  encounter newer isolated dumbbells going upwards, which occasionally results into the formation of a nested pair of dumbbells (which we call as dyads).
These dyads have two varieties, namely,
right handed dyads and  left-handed dyads (see left panel of Fig.\ref{fig:dyads_confg_data_comb}). The left and right handed dyads  are mirror images of each other.
Occasionally one obtains a transient structure like the pair marked
(NC) in Fig.\ref{fig:dyads_confg_data_comb} (left panel). This pair is not nested, and (consequently)
it is observed to be unstable. It is not chiral, being its own mirror image.

\item [$(iv)$] Right handed dyads with a small arbitrary initial heading  are observed to gradually change their heading to a particular value  $\theta_s >0$ and then maintain that heading but for minor fluctuations; see top panel of Fig. \ref{fig:dyads_a} and  Fig. \ref{fig:dyads}(a) and (b).  Similarly,  left handed dyads with a small arbitrary initial  heading  are observed  to gradually change their heading to $-\theta_s$ and then maintain that heading but for minor fluctuations; see bottom panel of Fig.\ref{fig:dyads_a} and  Fig. \ref{fig:dyads}(c) and (d). For the example shown in Fig.\ref{fig:dyads_confg_data_comb},   $\theta_s \approx 15^\circ$.
For a fixed radius of the balls, the steady state heading $\theta_s$ of the dyad decreases with increasing dumbbell lengths $\ell$ (see Fig.\ref{fig:length_sat}). 
 
\item[$(v)$] The qualitative features of the trajectories of a dyad are quite similar to those for a single dumbbell. When being carried up in $\varphi$ by the rotation of the cylinder, the two dumbbells touch each other and the pair moves up as a composite  object. As rolling is suppressed when objects are in contact, and as sliding
friction is stronger than rolling friction, the maximum angle $\varphi$ to which a dyad is carried up is greater that that for a single dumbbell.  The experimentally observed values of this angle are given in the right half  Fig.\ref{fig:s-curves} of the Appendix \ref{appn_friction}.
On reaching the maximum  value of $\varphi$ ,  the lower dumbbell in the pair breaks away from the top one by beginning to roll. It rolls down along a geodesic at an angle $\theta$ to the 
meridians, where $\theta$ is its heading. The higher dumbbell of the pair, whose heading is approximately the same
as that of the lower dumbbell, then follows the lower one along a nearby geodesic, till it comes to
a stop close to the first dumbbell, thus retaining the dyad structure ( \ref{fig:dyads_confg_data_comb}).
In this process the two dumbbells play the game of repeated rolling away and catching up. This behavior is similar to that for rolling spheres which was described in Kumar. et. al.   \cite{kumar2015granular}.

Given the continuous closeness of the two dumbbells in a dyad observed above, it is useful to 
attach a single position as well as heading to a dyad as a whole. The position 
$(\varphi, \eta)$ and the heading $\theta$ for a dyad will respectively mean the average of the positions and 
of the headings of the two dumbbells. Note that
the angle $\theta$ is well-defined only up to the addition of an integral multiple of  $\pi$.

\item [$(vi)$] The right handed dyads move up in $\eta$ till they reach the top of the cylinder, and then they fall out. The left handed dyads go to the bottom of the cylinder, where they break apart. 
As described above in (iii), new right handed dyads keep getting formed.  Our observations for our chosen 
experimental realization 
show  that  from a bunch of $60$ dumbbells that is initially placed in the bottom of the tilted rotating cylinder, about $40$ dumbbells exit the cylinder in 20 minutes, by forming right-handed dyads. 


\end{enumerate}

\section{Explanation of the observed phenomena for a single dumbbell}

A dumbbell experiences a different amount of friction for motion along its axis (sliding) and perpendicular to its axis (rolling), with the sliding friction being significantly higher than the rolling friction. 
This results in a `keel effect' in analogy with a boat in water that experiences very different resistances to moving forward as against sideways, while a raft -- lacking a keel -- does not show this behaviour. 
A detailed analysis of the frictional behaviour of a single dumbbell is given in Appendix \ref{appn_friction},
which includes the precise formulas as well as experimental observed values which validate the 
qualitative description given below.

A dumbbell placed in a tilted rotating cylinder at $\varphi =0$ with a small value of the heading
$\theta$ will get carried upwards along $\varphi$ by the rotation of the cylinder. 
Thus, its underlying part of the cylinder keeps getting steeper.
When $\varphi$ exceeds a certain value, the force of gravity overcomes the resistance of static friction.
Consequently, the dumbbell begins to roll down along the geodesic with constant $\theta$, till it reaches a lower value of $\varphi$ where it comes to a stop. This process gets iterated, 
leading to a zigzag trajectory.  Such trajectories for positive and negative values of $\theta$ are shown in 
respectively blue and green in Fig.\ref{fig:coordinates}, assuming that $\theta$ has remained constant. 

This auto toggling between stationary and rolling state allows
a dumbbell with a small positive value of the heading $\theta$ to move up the energy ladder in a sustained manner, provided it maintains a constant positive heading. However, the value of $\theta$ 
is not stable, and tends to zero as time passes for reasons of energetics that we now explain.
(The torque on a dumbbell, which leads to this change of heading, is discussed  
in Appendix \ref{appn_friction}.)

For this, it is convenient to regard a cylinder tilted at an angle $\alpha$ 
as a cylinder which is kept horizontal, in 
which an object of mass $ m $ placed on the surface is subjected to a body force 
in the direction $-\eta$ of magnitude $mg \sin\alpha$.   

A single dumbbell lying on the cylinder 
defines a point of the 
the space $M$ which has coordinates $(\varphi, \eta, \theta)$ where $(\varphi, \eta)$ describe  
the location of the center of mass of the dumbbell on the surface of the cylinder, and $\theta$ is a circular coordinate with period $\pi$
which describes the angle from a meridian line ($\varphi = const.$) to the axis of the dumbbell.
Geometrically, $M = \Omega\times S^1$ (where $S^1$ denotes a circle of circumference $\pi$) is the space of {\it apparent configurations} of a single dumbbell. This is a manifold of dimension three.

A dumbbell placed in a horizontal cylinder has a unique minimum potential
energy, which is achieved for $\varphi = 0$ and $\theta = 0$ (while $\eta$ may be 
arbitrary). The subset $M_{min}$ of $M$  defined by the 
simultaneous equalities $\varphi = 0$ and $\theta = 0$
is an attractive minimum locus. Thus,
slippages brought by noise lead to any dumbbell placed arbitrarily to move towards
$M_{min}$. When the heading $\theta$ is zero, the dumbbell rolls in place. 
Now suppose that we apply a body force which corresponds to a small value of $\alpha$. 
For a dumbbell with a small enough initial value of $|\theta|$, the 
effect of the body force does not interfere with this behaviour (where
$\theta$ gradually becomes $0$), but it 
tends to lower $\eta$ because of occasional slippages. 
If $|\theta|$ is large, then
the body force makes the dumbbell roll in a direction which lowers $\eta$. 
This explains the observed behaviour described above.

\section{The configuration space $D$ for a pair of dumbbells}

As seen above, a single dumbbell lying on the cylinder 
defines a point of the space $M = \Omega\times S^1$ of dimension three.
A similar geometric description for a pair of dumbbells begins with a point of the 
product space $M\times M$, with coordinates $(\varphi_1,\eta_1,\theta_1,\varphi_2,\eta_2,\theta_2)$
that describe the two dumbbells. However, not all points of  $M\times M$ are accessible, as 
the dumbbells cannot mutually interpenetrate. This results in unilateral constraints, which carve
out a subspace $D $ in  $M\times M$ as the actual configuration space for a pair. It turns out that
$D$ is not a manifold, but it is locally a polyhedron in curvilinear coordinates, having 
boundaries and corners. We now physically describe this curvilinear polyhedron in terms of the relative
placement of the two dumbbells. We will also physically describe the corners of $D$ and their interconnects in terms of allowed relative movements of the dumbbells.
Some of the basics of curvilinear polyhedra in manifolds are recalled in Appendix \ref{appn_polyhedron}. In this section, 
we describe the relevant geometry physically.

For simplicity, we will assume that the rods of the dumbbells are long enough so that 
a single ball of one dumbbell cannot simultaneously
touch both the balls of the other dumbbell.

\begin{figure}[t]
	\centering
	\includegraphics[width=1\linewidth]{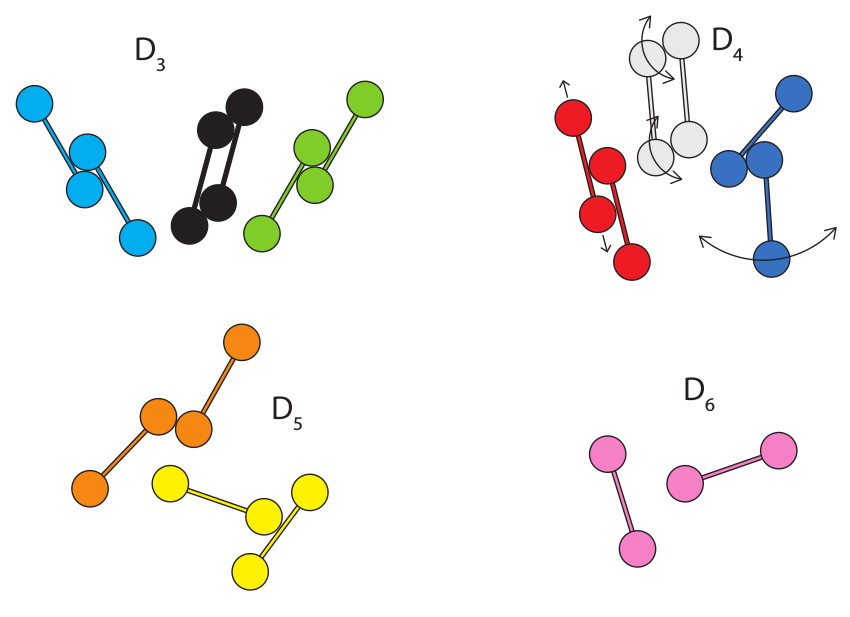}
	\caption{Dumbbell pairs corresponding to points on different faces of the polyhedron $D$. The blue dyad in $D_3$ is in $\Delta_R$,  the green dyad in $D_3$ is in $\Delta_L$ and the black dyad in $D_3$ is in $\mathcal{E}_L$. The three pairs shown in the $D_4$ portion of the figure, together with the mirror images of the red pair and the blue pair, give examples of points on all five components of $D_4$. Note that
		the gray pair in the $D_4$ portion of Fig.\ref{fig:geometry_of_dyads} and its the mirror image lie in the 
		same component of $D_4$. The dumbbell configurations which correspond to points in the manifold $D_5$ have no chirality, and there are exactly three connected components. One of these corresponds to the orange pair in the $D_5$ portion of the figure, and 
		two other components correspond to the yellow pair in the $D_5$ portion of the figure
		(these two components are related to each other by interchanging the two dumbbells).} 
	\label{fig:geometry_of_dyads}
\end{figure}

The set $D$ is six dimensional, and it is naturally partitioned into subsets 
$D_i$ for $i=3,4,5,6$ which have direct physical descriptions (see Fig.\ref{fig:geometry_of_dyads}).
A dummbell pair where the dumbbells are not touching each other defines a point of $D_6$. If the pair has a one point contact then the corresponding point of $M\times M$ is in $ D_5 $ and for a two point contact the corresponding point is in $ D_4 $. Points of $ D_3 $ represent dumbbell pairs with at least three points of contact.  
As will be apparent, the sets $D_i$ have dimension $i$ for $i = 3,4,5,6$. These are the faces of the polyhedron $D$.
It is significant that $D_3$ is the smallest dimensional nonempty face (that is, the faces 
$D_0$,  $D_1$ and $D_2$ are empty). In other words, $D_3$ is the sharpest corner of the polyhedron 
$D$, being the smallest dimensional face.

If we have a pair of dumbbells which do not touch, then if each dumbbell is independepently perturbed by
small enough amount, then they continue not to touch. This shows that $D_6$ is an open set, and so it has 
dimension $6$. A pair which corresponds to a point of the set $D_5$ has two kinds of realizations
depending on whether  
the single contact point of the two dumbbells lies on a ball of each dumbbell, or lies on the ball
of one dumbbell and the rod of the other dumbbell (see Fig.\ref{fig:geometry_of_dyads}).  In a small neighbourhood in $M\times M$, one can
unambiguously label the centres of the balls of the first dumbbell as $A_1$, $A_2$ and of the second dumbbell as $B_1,B_2$. 
Suppose we have a point of $D_5$ for which the ball with centre $A_i$ touches the ball with centre 
$B_j$. Then in a small enough neighbourhood  
of this point in $M\times M$, the set $D$ is described by the inequality
$$d(A_i,B_j) \ge 2r$$
where $r$ is the radius of the balls, and $d$ is the distance function.
The portion of $D_5$ is locally described by the corresponding equality 
$d(A_i,\beta) = 2r$ while the portion of $D_6$ is locally described by the corresponding strict inequality 
$d(A_i,\beta) > 2r$.
If we have a point of $D_5$ for which the ball with center $A_i$ touches the rod of the other dumbbell. Then in a small enough neighbourhood  of this point in $M\times M$, the set $D$ is described by the inequality
$$d(A_i,\beta) \ge r +t$$
where $\beta$ is the axial line of the second dumbbell, and $t$ is the radius of the rod. 
The portion of $D_5$ is locally described by the corresponding equality 
$d(A_i,\beta) = r +t$ while the portion of $D_6$ is locally described by the corresponding strict inequality 
$d(A_i,\beta) > r +t$.
A similar inequality
$d(\alpha, B_j) \ge r+t$ (and the corresponding equality and strict inequality) works in a neighbourhood of a point of $D_5$ for which the rod of the first dumbbell
touches a ball of the second dumbbell, where $\alpha$ is the axial line of the first dumbbell.

It can be seen that points of $D_4$ correspond to dumbbell pairs which have five kinds of realizations, some of which are shown in Fig.\ref{fig:geometry_of_dyads}. The set $D_4$ can be locally described in a small enough neighbourhood of any of 
these 5 kinds of points by suitable inequalities in terms of distances. For example, if 
a point of $D_4$ is realized by a pair such as the red pair in the $D_4$ portion of Fig.\ref{fig:geometry_of_dyads},  then 
the simultaneous inequalities 
\begin{eqnarray*}
d(\alpha, B_1) &\ge &r+t\\
d(A_2, \beta)& \ge & r+t
\end{eqnarray*}
define $D$ in a small enough neighbourhood. The corresponding
equalities define $D_4$ locally. Making both inequalities strict defines $D_6$ locally,
while making exacly one of these into an equality defines $D_5$ locally.

The set $D_3$ is of special interest to us. 
The points of $D_3$ have two kinds of realizations, leading to
a partition $D_3 = \Delta \cup \mathcal{E}$. Points of $\Delta$ are realized by pairs 
which have a three point contact. These appear like the blue or green pair
in the $D_3$ portion of Fig.\ref{fig:geometry_of_dyads}. Points of $\mathcal{E}$ correspond to pairs which 
have a four point contact. These appear like the black 
pair in the $D_3$ portion of Fig.\ref{fig:geometry_of_dyads}, or like its mirror image.

Suppose we have a point of $D_3$ for which the balls with center $A_2$ and $B_1$ touch each other and 
also touch the rod of the other dumbbell (see the pair $P_1$ in Fig.\ref{fig:half_spaces_new1}).
Note that this is a point of $\Delta$. 
In a small neighbourhood of this point in $M\times M$, the portion of $D$ is
defined by the inequalities
\begin{eqnarray*}
	d(A_2,\beta) &\ge & r +t,\\ 
	d(\alpha, B_1) &\ge & r+t,\mbox{ and}\\
	d(A_2,B_1) &\ge  & 2r.
\end{eqnarray*} 
As before, by replacing various inequalities by strict inequalities or equalities,  
one obtains the portions of $D_6$, $D_5$, $D_4$ and $D_3$ in this neighbourhood.

Suppose we have a point of $D_3$ for which the balls with centers $A_1$ and $B_1$ touch each other,
the balls with centers $A_2$  and $B_2$ touch each other, the balls with centers $A_2$ and $B_1$ 
also touch the rod of the other dumbbell (see the pair $P_2$ in Fig.\ref{fig:half_spaces_new1}). 
Note that this is a point of $\mathcal{E}$. 
Then in a small neighbourhood of this point in $M\times M$, the portion of $D$ is
defined by the inequalities 
\begin{eqnarray*}
	d(A_2,\beta) &\ge & r +t,\\ 
	d(\alpha, B_1) &\ge & r+t,\\
	d(A_1,B_1) &\ge & 2r, \mbox{ and}\\
	d(A_2,B_2) &\ge & 2r.	
\end{eqnarray*} 
The corresponding four equalities (but not the original inequalities) 
are overdetermined by one, so the set $\mathcal{E}$ is $3$ dimensional.

Taking chirality into consideration, 
we get a finer partition 
$$D_3 = \Delta_L \cup \Delta_R \cup \mathcal{E}_L \cup \mathcal{E}_R$$
where $\Delta_L \cup \Delta_R = \Delta$ and $\mathcal{E}_L\cup \mathcal{E}_R  = \mathcal{E}$, and 
where the subscripts $L$ and $R$ denote the left or right chirality of the pair.
For example, the blue pair in Fig.\ref{fig:geometry_of_dyads} defines a point of $\Delta_R$ and the green 
pair in Fig.\ref{fig:geometry_of_dyads} defines a point of $\Delta_L$. The black pair in Fig.\ref{fig:geometry_of_dyads} defines a point of $\mathcal{E}_L$. 
The concept of chirality can be made precise as follows.
For any pair of dumbbells in $D_3$, there is a unique pair of centers $A_i$, $B_j$ 
of balls which are furthest apart from each other. Let $A_p$ and $B_q$ be the remaining centers.
The dumbbell pair is left-handed (respectively, right handed) if and only if 
the pair of vectors $( \overrightarrow{ A_iB_j}, \overrightarrow{A_pB_q})$
is left-handed (respectively, right handed). The subsets $\Delta_L$ and $\mathcal{E}_L$ of $\Delta$ and $\mathcal{E}$ respectively  consist of the left-handed pairs  while the subsets   $\Delta_R$ and $\mathcal{E}_R$ of $\Delta$ and $R$ respectively  consist of the right-handed pairs.

Each of the components $\Delta_L$, $\Delta_R$, $\mathcal{E}_L$ and $\mathcal{E}_R$ 
consists of all points of $M\times M$ that can be obtained by translating or rotating 
the dumbbell pair which defines any particular representative point of that component. This 
gives an identification of each of $\Delta_L$, $\Delta_R$, $\mathcal{E}_L$ and $\mathcal{E}_R$ with $M$.

Though $D$ is six dimensional, because $3$ of those dimensions are free (as can be seen by 
translating or rotating the dumbbell pair representing any point of $D$), 
one can  give a sense of how $D$ appears (at least locally) by means of a figure in $6-3 = 3$ dimensions.
When the $3$ free dimensions in $D$ are suppressed, the 
region $D$ locally appears like the exterior of the solid object depicted in Fig.\ref{fig:half_spaces_new1}(c). The compact solid object locally represents interpenetrating
pairs. 
The corners and edges of this solid body are {\it inverted}, that is, they point inwards.
Note that $D_3$ is locally represented by the corner points of this body, while $D_4$ locally 
appears as its edges. The $2$-dimensional faces locally represent $D_5$. The region $D_6$ is locally
represented by the complement (exterior) of the body. Note that the corners and edges of the complement point
outwards, making the complement locally convex in a curvilinear sense. 

The parts Fig.\ref{fig:half_spaces_new1}(a) and Fig.\ref{fig:half_spaces_new1}(b) show how to move the two dumbbells relative to each other, while 
retaining contact, so as to go from a point of one of the components 
$\Delta_L$, $\Delta_R$, $\mathcal{E}_L$ and $\mathcal{E}_R$  of $D_3$ to another component, while remaining within $D_4$.
The various motions in Fig.\ref{fig:half_spaces_new1}(a) and Fig.\ref{fig:half_spaces_new1}(b) are color coded, and the corresponding paths in $D$ 
are marked by the same color in Fig.\ref{fig:half_spaces_new1}(c).

The polyhedron $D$ is locally convex in a curvilinear sense (as formally defined in Appendix \ref{appn_polyhedron}). 
The significance of this for the energetics and the stability of a dumbbell pair will become clear later.

\section{Energetics and entrapment over $D$}
For a pair of dumbbells in a horizontal cylinder, we now study the potential energy function $E$ on the space $D$. 
 For simplicity,
 we will neglect  the weight of the  rods of the dumbbells. 
The function $E$ is the sum of the  individual potential energies of the dumbbells.
The absolute minimum value for $E$ is achieved when both dumbbells lie along the lowest meridian of the 
cylinder.
In coordinate terms, the corresponding point $(\varphi_1,\eta_1,\theta_1,\varphi_2,\eta_2,\theta_2)$ of
$D$ satisfies $\varphi_1 = \varphi_2 =0$ and $\theta_1 = \theta_2 = 0$.  
Such points lie in $D_6$ when the dumbbells do not touch each other, or in $D_5$ as a limiting case when the dumbbells
touch at one point. They form a subset $G\subset M\times M$. As the dumbbells cannot inter-penetrate, $G$ has two components, respectively  
defined by the conditions $\eta_1 - \eta_2 \ge \ell + 2r$ or 
$\eta_2 - \eta_1 \ge \ell + 2r$. 

Besides the global minimum for $E$, which is attained on $ G $, it turns out that there is another local minimum value for $E$, which is attained along 
a locus $\delta$ which lies inside $\Delta$ in $D_3$. Chirality  of the pairs gives a decomposition
$\delta = \delta_L \cup \delta_R$ where $\delta_L \subset \Delta_L$ and 
$\delta_R \subset \Delta_R$. It is shown later that $G$ and $\delta$ are the only local minimal
energy loci.

In terms of distance functions and coordinates, the sets $\delta_L$ and $\delta_R$ are
defined as follows. 
A point $(\varphi_1, \eta_1, \theta_1,\, \varphi_2, \eta_2, \theta_2)$ of $\Delta_L$ 
lies in $\delta_L$ if and only if 
$$
\varphi_1 + \varphi_2 =0 ~\rm{ and }~ 
	\theta_1 = \theta_2 = -\arctan(r/\ell).
$$
The subset $\delta_R \subset \Delta_R$ is analogously defined by
$$
\varphi_1 + \varphi_2 =0  ~\rm{ and }~ 
\theta_1 = \theta_2 = \arctan(r/\ell).
$$
The sets $\delta_L$ and $\delta_R$ are disjoint closed submanifolds of $M\times M$,
each isomorphic to the real half line with coordinate $\eta = (\eta_1+\eta_2)/2$ (the average $\eta$).
Sample pairs in the sets $\delta_L$ and $\delta_R$ are shown in the right panel of  Fig.\ref{Fig: nested_2}. 

\begin{figure*}[htb]
	\centering
	\includegraphics[width=1\linewidth]{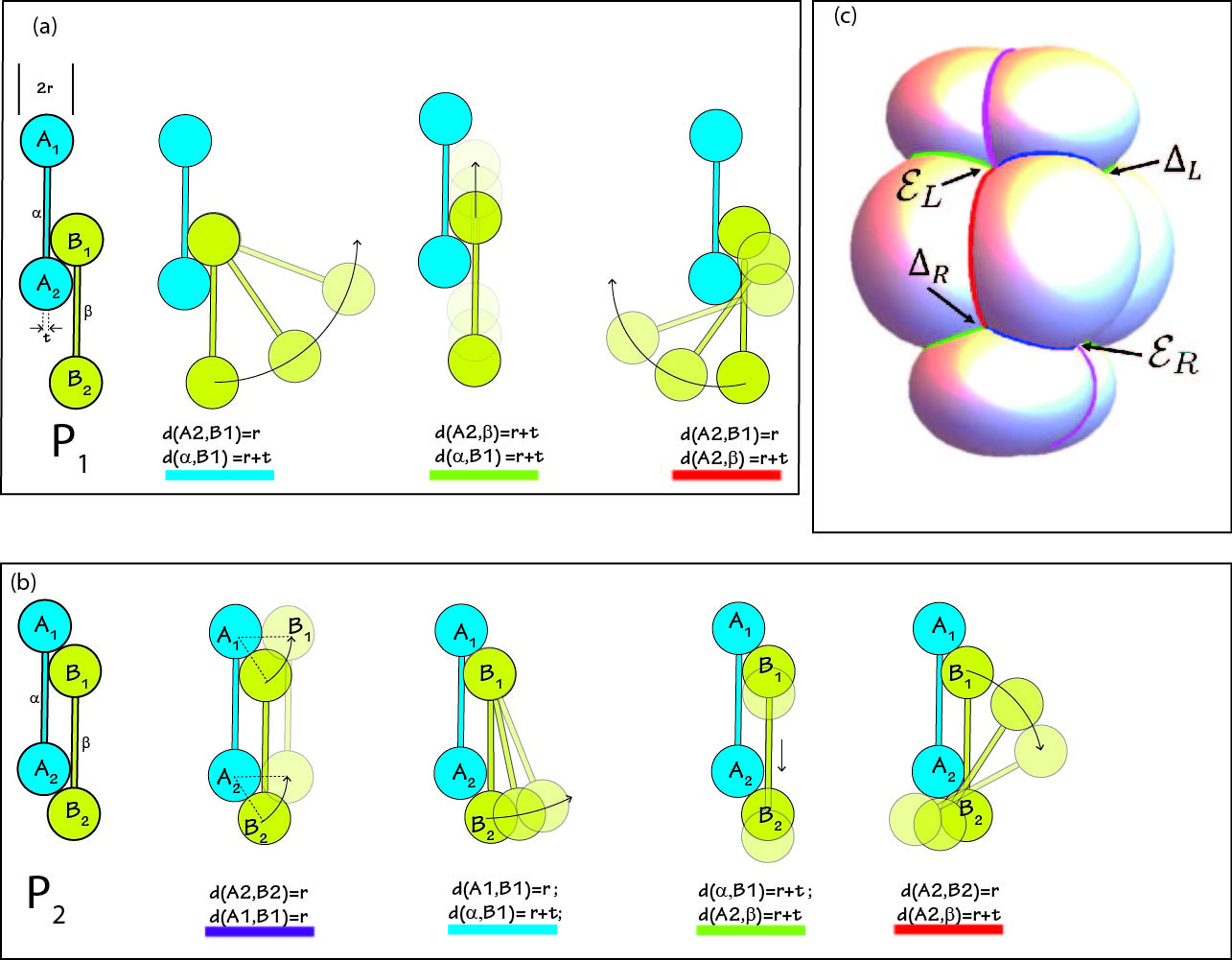}
	\caption{ (a) The left-most dyad defines a point in $\Delta_R$. The next three figures shows the  allowed half motions  of the  dyad along which certain distance functions are held constant. 
	Each motion is given a color code.	
		 (b) The left-most dyad corresponds to a point of $\mathcal{E}_R$. The next four figures shows the  allowed half motions  of the dyad. Besides the 3 motions from (a), there is a new motion coded
		 purple. (c) The  region $D$  locally looks like the product of $\R^3$  and the exterior of the depicted translucent solid object. The  sets $\Delta_L$, $\Delta_R$, $\mathcal{E}_L$ and $\mathcal{E}_R$ correspond to the corners so marked. The paths between the corners, coming from the motions listed in (a) and (b), 
		are marked in corresponding colors.} 
	\label{fig:half_spaces_new1}
\end{figure*}

The proof of the fact that $\delta$ is a local minimal set of configurations for energy 
is a consequence of the geometry of $D$, as treated in the Appendix \ref{appn_enrgy}. As
$\delta$ is a subset of the corner $D_3$, the continued dissipation by inelastic collisions at the boundary,
which physically correspond to the two dumbbells colliding against each other, 
entraps the configuration point  into the set $ \delta $.  
In fact,  as the total derivative of $E$ does not
vanish over $\delta$, minimization of energy leads to a stronger entrapment of the configuration
point than what more common `U-shaped' or `half-U-shaped' potentials would give, 
as explained in Appendix \ref{appn_enrgy}. In terms of the Appendix \ref{appn_enrgy}, 
the entrapment which organizes a pair of dumbbells into a dyad in $\delta$ 
is of `half-V' (triangular) potential type in one degree of freedom (the one which corresponds to the relative motion 
depicted in the 3rd panel of Fig.\ref{fig:half_spaces_new1}(a)), of `half-U' (half harmonic) potential type in two degrees of freedom (depicted in the 2nd and 4th panels of Fig.\ref{fig:half_spaces_new1}(a)), of `full-U' (harmonic) potential type in two degrees of freedom (where the 
pair moves together by changing $\theta$ or $\varphi$),
while the motion is free (not counting friction) along the $1$-dimensional subset $\delta$
as the energy is independent of $\eta$. 

In the next section, we discuss the motion of dumbbells when the cylinder undergoes a noisy rotation. 
If the level of noise and the angular speed of rotation are small enough then the 
organization of the dumbbell pairs survives their destabilizing influence, showing the strength of the entrapment.


\section{Motion of a dumbbell dyad}

\subsection{Dyads and the subset $T\subset M\times M$}

\begin{figure}[t]
	\centering
	\includegraphics[width=1\linewidth]{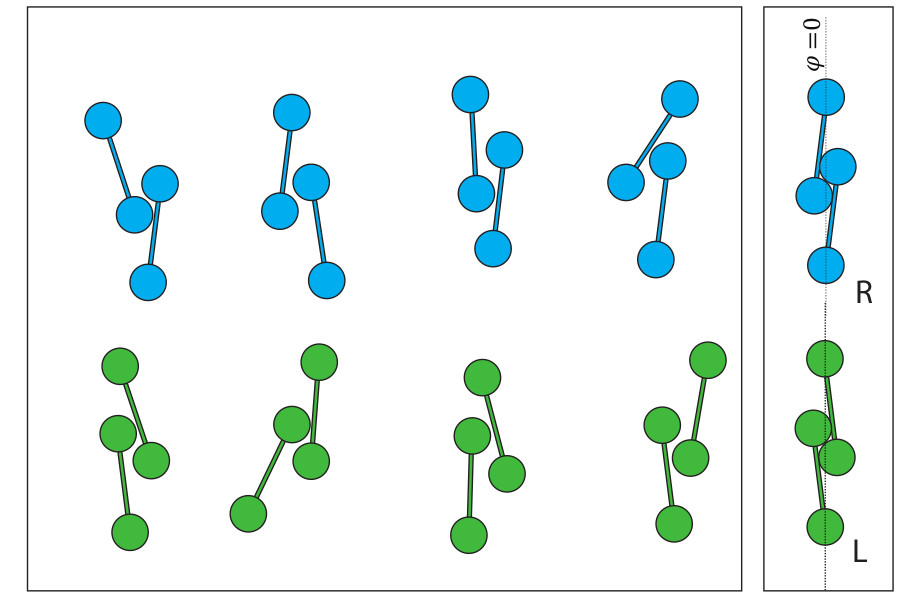}
	\caption{Representative configurations of dumbbell dyads in a neighbourhood of $\delta_R$ 
	(respectively, $\delta_L$)	
	are shown in blue (respectively, in green) in the left panel. The right panel shows
	 energy minimizing dyads of either chirality, which 
	 lie in $\delta_R$ and $\delta_L$, respectively.}
	\label{Fig: nested_2}
\end{figure}

To make the notion of a dyad more precise, we first identify a subset $T \subset M\times M$ of these.
By definition, $T$ consist of all  pairs of apparent configurations 
such that the two dumbbells are parallel to each other, and one ball of
each dumbbell touches the rod of the other dumbbell. 
Note that $T$ is a closed subset of $M\times M$, and it is in fact the union of all of $D_3$ with exactly 
$2$ of the $5$ connected components of $D_4$, namely, those which
correspond to the red pair in the $D_4$ portion of Fig. \ref{fig:geometry_of_dyads} and its mirror image.

One may say in general that a dyad is a point in $M\times M$ which is in a small neighbourhood of $T$ with respect to the natural metric on $M\times M$ induced from that on $M$. 
One possible choice of such a neighborhood (for the sake of definiteness) is  the open subset $U\subset M\times M$ defined by the inequalities
\begin{eqnarray*}
 max_i\, d(A_i,\beta) &< & 2r,\\
 max_j\, d(\alpha, B_j) &< & 2r,\\
 max_{i,j}\,d(A_i,B_j) &< & 2\sqrt{ \ell^2 + r^2}.
\end{eqnarray*}
Every dumbbell pair inside this neighbourhood has a chirality. More generally, 
keeping the width of the neighbourhood small ensures that the pair of dumbbells remains
nested, and so it has a chirality. 

We have already explained that $G$ and $\delta$ are loci of local minima for energy.
Inspection shows that for any point of $D$ outside $G\cup \delta$, there exists a nearby 
point in $D$ where the total energy is lower. This shows that the subset  
	$G\cup \delta$ is the entire set of all points of local minima for the total energy function
	on $M\times M$ for a pair of dumbbells in a horizontal cylinder.

\subsection{Pairs in a rotating horizontal cylinder.} 
We will now consider the case of a rotating horizontal cylinder. The following discussion applies
when the level of noise as well as the angular speed of rotation are both sufficiently small.
In such a cylinder, energy minimization 
takes the point representing a dumbbell pair to either a small neighborhood of 
$\delta$ or to a small neighborhood  of $G$.  Representative configurations of the dumbbell pairs 
close to $\delta_L$ (in blue) and $\delta_R$ (in green) are shown in the left panel of Fig.\ref{Fig: nested_2}.

If the point representing the pair is close to $\delta_L$ (respectively to $\delta_R$), then in physical
terms, the dumbbell pair is a left handed (respectively right handed) dyad with a small negative (respectively positive)
heading, of the order of $ \arctan(r/\ell)$. Therefore, 
its subsequent motion is a zigzag as explained in 3.2.(v) above.
On the other hand, if 
the point representing the pair is close to $G$, then both the dumbbells roll in place,
and the point representing the pair roughly remains stationary in $M\times M$.

\subsection{Pairs in a tilted rotating  cylinder.} 
As before, we can view a tilted cylinder as a horizontal cylinder together
with a sideways body force field $f_H$ acting in the direction $-\eta$. This 
body force causes occasional slippages of the pair in the $-\eta$ direction,
both in the case of pairs moving on zigzag path or pairs rolling in place.
 The effect of these slippages in terms of movement in the $-\eta$ direction 
is not  too large if the body force is not too large.
The body force causes a change of heading for pairs in $\delta$, which again is
small when $f_H$ is small. In fact, this change of heading is negligible for the 
actual experimental parameters (see Fig.\ref{fig:length_sat}).
 Hence the headings 
remain non-zero positive (or negative), close to $\pm \arctan(r/\ell)$ with small fluctuations.
Hence a pair in $G$, which would have essentially rolled in place in the absence of 
a body force, now moves via slippages in the $-\eta$ direction, eventually reaching the bottom. 
A pair near $\delta_L$ would have moved in a zigzag path reducing $\eta$
in the absence of a body force. This movement tin the $-\eta$ direction is enhanced
by the slippages, and such a pair also goes to the bottom of the cylinder.

The case of interest are pairs which define a point close to $\delta_R$. Such a pair
would have moved in a zigzag path increasing $\eta$
in the absence of a body force. 
If the noise and the body force are not too large, this zigzag motion which increases the value of $\eta$ is 
not entirely negated by the occasional slippages in the $-\eta$ direction. 
Hence such a pair goes to the top of the cylinder.

It should be noted that after the value of $\theta$ for a right handed dyad arrives close to $\theta_s$, 
the variation in $\varphi$ during its subsequent zigzag trajectory is about $5^{\rm o}$,
and $\theta$ continues to remain within $\theta_s\pm 2^{\rm o}$. This means
that the dyad stays close to $\delta_R$ during its zigzag trajectory, which re-enforces the stability
of the heading and of the dyad formation. The zigzag trajectory moves upwards along $\eta$, which parameterizes
the half-line $\delta_R$, and the arms of the zigzag do not extend very far from the attractive set $\delta_R$
for the energy function $E$ on $D$.

\subsection{Formation of dyads}
We now continue with a tilted rotating  cylinder, again assuming that the noise and the tilt are not too large.
Up to now, we have analysed the motion of either a single dumbbell or the motion of a pair of dumbbells placed 
in the cylinder. Let us now consider the general case where a number of dumbbells are placed in the cylinder.
An individual dumbbell with positive slope in the appropriate range will rise in $\eta$ by a zigzag path as explained earlier. But eventually, such the heading of such a dumbbell goes to zero, and 
the dumbbell starts coming downwards. This descending dumbbell may encounter another dumbbell
which is rising, and these two dumbbells may collide to become approximately parallel, and come to define a point of $M\times M$ which is in the attractive basin around the local minimum set $\delta$. 
This results in the formation of dyads. 

When a left handed pair reaches the bottom, it breaks apart. 
These get added to the dumbbells at the bottom, along with single dumbbells
which come all the way down. Such dumbbells may become a part of new
right handed dyads by the process described above. 

This is how ever new right handed dyads keep getting formed. Such a dyad, unless it gets obstructed 
by other dumbbells, rises to the top of the cylinder and falls off.

\subsection{Chiral sorting}
An interesting outcome of the dependence of the direction of motion on the chirality of a dyad is that it leads a sorting of  dyads by their chirality.  It is noteworthy that the underlying mechanism for this sorting is achiral in a strong sense, more precisely,
a rotating horizontal cylinder which is infinite in both directions is a rigid body 
with an achiral motion.
de Gennes has given another instance of an achiral mechanism for sorting  (see \cite{de1999mechanical}). 
Such achiral mechanisms may be contrasted with the more common `hand-in-glove' approach to sorting of chiral objects, which relies on a chiral environment (such as parallel electric and magnetic fields as first proposed by P. Curie \cite{Curie}),  or on the  initial provision of a model chiral object as a template for sorting.

\section{Conclusion }  

We have seen by a detailed analysis of dumbbells placed in a tilted rotating  cylinder how the major qualitative
aspects of the behaviour are explained in terms of an underlying polyhedral geometry. This included the formation
and stability of dyads of two chiralities, and their sustained locomotion. 

The unilateral constraints which gave rise to the polyhedron $D$ arose from the mutual non-interpenetrability of
dumbbells. These constraints were expressible by inequalities which take the form $ f_j(q_i) \ge 0 $ in terms of  the generalized position coordinates $ q_i  $ of the dumbbells.
More generally, there can exist physical situations (e.g. location dependent 
speed limits in traffic rules) where the unilateral constraints involve generalized 
velocities and take the form $ f_j(q_i,\dot{q_i})\ge 0 $. Under appropriate assumptions, these will carve out polyhedra in phase spaces of the physical systems. Again, interesting physics
can be expected to take place in a neighbourhood of the corners of these polyhedra. 

One should note that the polyhedron $D$ in this paper was locally convex in a curvilinear sense. 
Such locally convex corners can lead to entrapment via an optimization mechanism explained in Appendix \ref{appn_enrgy} 
or via its suitable generalizations to more general types of corners. 
Other physical systems lead to polyhedra which may not be locally convex, and which may in fact admit `locally concave' 
corners, which result in bifurcations in the systemic evolution. A simple example of this phenomenon, which involves
both locally convex and locally concave corners, leading to entrapments and bifurcations,  
is given by the geometry and energetics of a pinball (bagatell) machine. 

For simplicity, we have only considered the differential category so far. But
given that many physical systems correspond to spaces defined by algebraic equations, one should expect
that `semi-algebraic sets' (loci defined by inequalities involving algebraic functions on
varieties) occur in place of curvilinear polyhedra, allowing more general singularities 
than corners (e.g., cusps) to occur.  

The prevalence of the English expressions
`to corner' and `to drive a wedge' for describing the processes of entrapment or bifurcation is 
actually a recognition of the role of polyhedra and their corners
which underlie diverse phenomena. 
Given the ubiquity of unilateral constraints (e.g., steric hinderence in molecules \cite{weinhold2001chemistry}) 
and also of entrapments (e.g., folded states of proteins \cite{ramachandran1963stereochemistry}) and bifurcations in the physical world,
we expect that it is worth looking for explanations of such phenomena based on curvilinear polyhedra in 
configuration or phase spaces, and dynamics near their corners.

\appendix

\section*{Appendix}
\renewcommand{\thesubsection}{\Alph{subsection}}

\subsection{Frictional behaviour of a single dumbbell.}

\subsubsection*{ The keel effect}\label{appn_friction}
\begin{figure}
	\centering
	\includegraphics[width=0.9\linewidth]{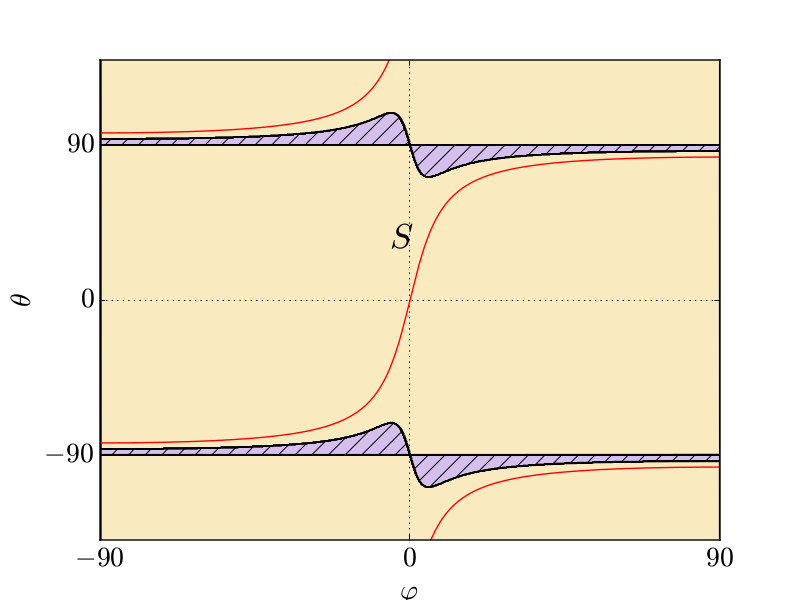}
	\caption{The $S$  curve is  marked as a red line. The  outside of the hatched region is the subset of the points in $N$ where the net torque tends to decrease $|\theta|$. Inside the hatched region the net torque tends to increase $|\theta|$ towards $ \pi/2 $.  This figure is plotted for $ \alpha=7^{\circ} .$}
	\label{fig:S_curve_torque}
\end{figure}

\begin{figure}[b]
	\centering
	\includegraphics[width=0.9\linewidth]{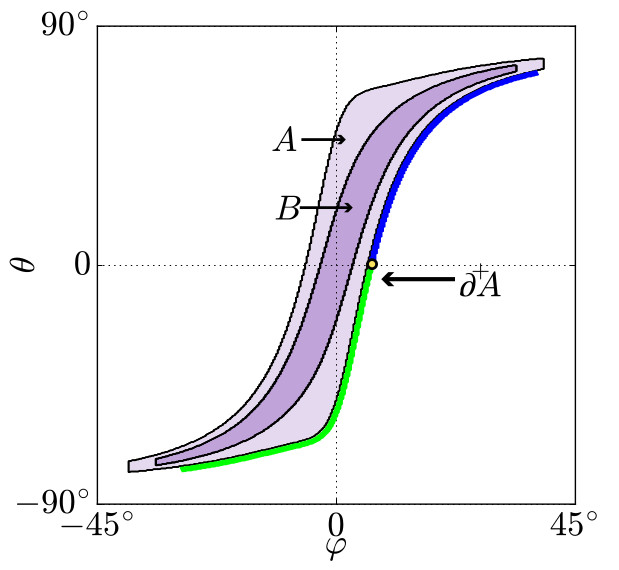}
	\caption{Schematic diagram of sets $A, \, B, \, \partial^+A $ of points defined in $N$ for a tilted cylinder. The blue part of $ \partial^{+} A $ corresponds to dumbbells with positive headings while the green part of $ \partial^{+} A $ corresponds to dumbbells with negative headings. }
	\label{fig:s-curve_schematic}
\end{figure}

\begin{figure*}
	\centering
	\includegraphics[width=1\linewidth]{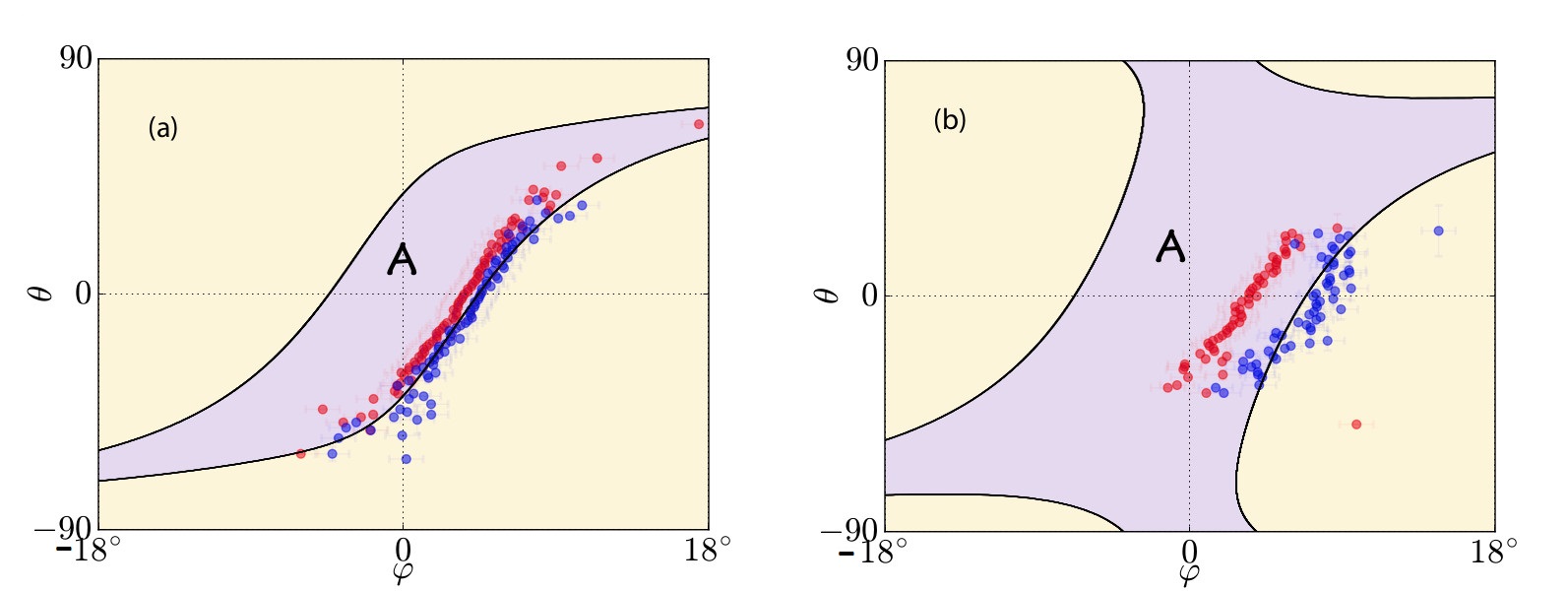}
	\caption{Measured values of  $\varphi_s$  as a function of $\theta$ are marked as blue points for single dumbbells in (a) and for dyads in  (b). At these values of $\varphi$ the dumbbells (dyads) that are being carried up by a tilted rotating cylinder with $ \alpha=7^{\circ} $  begin to slip or roll downwards. These points can be seen to lie approximately on the  eastern boundary of $A$. 	The red points  mark the   values of $\varphi$ at which moving dumbbells (or dyads) came to a rest with respect to the surface of the cylinder (which  happens in the region $B$ which is not marked in the figure). 
		The experimental data corresponds to the following parameters 
		$\mu_r^{stat}=0.1, \,\epsilon^{stat}=8^{\circ}$ (dumbbell) and $\mu_r^{stat}=0.2,\, \epsilon^{stat}=20^{\circ}$ (dyad).  
		The  physical parameters for the dumbbells are $\ell=16 \,\mbox{mm},\,  r=3 \,\mbox{mm}, \, t=1.5 \,\mbox{mm}$. }
	\label{fig:s-curves}
\end{figure*}

The frictional resistance to the onset of motion of a stationary dumbbell lying on a stationary substrate is captured by two dimensionless constants which we denote by 
$\mu^{stat}_s$ and $\mu^{stat}_r$, where $0 < \mu^{stat}_r < \mu^{stat}_s$. 
These two constants have the following operational definitions.
If a dumbbell lying stationary on a surface is subjected to a force $f_N$ normal to the surface and 
a force $f_r$ tangent to the surface in the direction perpendicular to the axis of the dumbbell,
then it starts moving (which will be mainly by rolling) provided $||f_r||/||f_N|| > \mu^{stat}_r$.
Here it is assumed that 
the left hand side is only slightly larger than the right hand side. 
Instead of the force $f_r$, if a force $f_s$ along the axis of the dumbbell is applied, then the dumbbell starts moving (which will be by sliding) provided $||f_s||/||f_N|| > \mu^{stat}_s$. 
In general, empirical observation shows the following (which is an idealized description that 
ignores mechanical noise and the statistical irregularities of the surfaces). 
Let  a force $f_T$ tangent to the surface be applied to the dumbbell, making an 
angle $\vartheta$ with the perpendicular direction  to the dumbbell (its direction of rolling). Suppose that $|\vartheta| < \pi/2 - \epsilon^{stat}$ where 
$$\sin\epsilon^{stat} = \mu^{stat}_r/\mu^{stat}_s.$$
Under the application of such  a force, the dumbbell begins to move (mainly by rolling) if 
$ ||f_T||/||f_N|| > \mu^{stat}_r/\cos \vartheta $. On the other hand, if 
$|\vartheta| > \pi/2 - \epsilon^{stat}$,
then the dumbbell begins to move (by a mixture of sliding and rolling) if 
$||f_T||/||f_N|| > \mu^{stat}_s$. 
Measurement shows that we have the values $\mu^{stat}_r \sim 0.1$, $\mu^{stat}_s \sim 0.3$
and $\epsilon^{stat} \sim 20^o$ for the dumbbells and the substrate (the glass cylinder) used in our experiment.

In the above described case, where $|\vartheta| < \pi/2 - \epsilon^{stat}$
and $ ||f_T||/||f_N|| > \mu^{stat}_r/\cos \vartheta $, observation shows that when a dumbbell
begins to move, it moves  by rolling along a geodesic trajectory on the surface 
which is perpendicular to the axis of the dumbbell, modulo fluctuations brought about by
mechanical noise and the statistical irregularities of the surfaces. Here we have assumed that 
the reciprocal of the mean curvature of the surface is everywhere significantly greater than the length of the dumbbell.

A dumbbell moving slowly on a stationary substrate can be brought to a halt by frictional forces. This phenomena of the cessation of motion is controlled by analogous coefficients $\mu^{dyn}_r$ and $\mu^{dyn}_s$ of friction in motion. These coefficients are considerably smaller than the corresponding coefficients $\mu^{stat}_r$ and $\mu^{stat}_s$ which control the onset of motion. Consequently, the frictional resistance offered by the surface to a dumbbell decreases as soon as it starts to roll. If we ignore the effects of inertia, then the condition for a rolling dumbbell to come to a halt is $ ||f_T||/||f_N|| < \mu^{dyn}_r/\cos \vartheta $ in terms of the notation used above.

\subsubsection*{ Dumbbells in a tilted stationary cylinder.} 
We now apply the above equations to determine when a stationary dumbbell starts rolling, and when a rolling dumbbell comes to a halt, in our experimental
setup. Here, the dumbbell is placed on the inside surface of a cylinder as described earlier,
and is subjected only to gravitational and frictional forces.
From the frictional properties of a dumbbell, described above, we can determine its
trajectory. Of course, this is an idealization which ignores the effect of noise and random slippages.

Our experimental parameters satisfy the following inequalities
\begin{equation}
\mu_r^{dyn} < \mu_r^{stat} < \tan \alpha < \mu_s^{dyn} <\mu_s^{stat}.	
\label{Eqn:parameters1}
\end{equation}
Physically, the inequalities $\mu_r^{dyn} < \mu_r^{stat}$ and $\mu_s^{dyn} <\mu_s^{stat}$ 
mean that dynamic friction is smaller than the corresponding kind of static friction. 
The inequalities 
$\mu_r^{dyn} < \mu_s^{dyn}$ and $\mu_r^{stat}  <\mu_s^{stat}$
mean that rolling is easier than sliding.	
The inequality 
$\mu_r^{stat} < \tan \alpha$ means that a stationary dumbbell placed at $\varphi =0$ with
$\theta = \pi/2$ begins to roll downwards (i.e., the slope of the cylinder is not too small).
The inequality  
$\tan \alpha < \mu_s^{dyn}$
means that a stationary dumbbell placed at $\varphi =0$ with
$\theta = 0$ will not slide downwards even when given a small nudge 
(i.e., the slope of the cylinder is not too large).
It is an empirical fact that both the inequalities 
$\mu_r^{stat} < \tan \alpha$ and $\tan \alpha < \mu_s^{dyn}$ can be simultaneously
satisfied when $\alpha$ lies in a certain nonempty interval.







%










Let $\varphi_s$ be the angle so defined that if any object made of the same material
as the dumbbell is placed at a point $(\varphi,\eta)$ on the cylinder, with $|\varphi|> \varphi_s$,
then the object begins to move. It can be seen that 
\begin{equation}
\varphi_s = \sin^{-1}
\left(\sqrt{\frac{ (\mu^{stat}_s)^2-\tan^2 \alpha} {1+(\mu^{stat}_s)^2}}\right)
\label{Eqn:varphi_s}
\end{equation}
The downward pointing unit vector field $-\partial/\partial z$ in the laboratory, when restricted to the surface of the cylinder, has an orthogonal decomposition $\partial/\partial z = F_T + F_N$ with $F_T$ tangent and $F_N$ normal to the surface. Recall that $F_T$ is given by Eqn.(\ref{Eqn:tangent_force}). Hence  $F_N$ has the magnitude
$$|F_N| =  \sqrt{1 - |F_T|^2} = \cos \alpha \cos \varphi.$$
As the gravitational force on a dumbbell is given by
$f = -mg \,\partial/\partial z$, we get $f_T = mg F_T$ and $f_N = mg F_N$ in terms of the vector fields $F_T$ and $F_N$, and so
$$\frac{f_T}{|f_N|} = \frac{F_T}{|F_N|} 
= -\tan\varphi \,R^{-1}\partial_{\varphi} -\frac{\tan\alpha}{\cos\varphi}\,\partial_{\eta} 
$$
If a dumbbell located at $(\varphi, \eta)$ has heading $\theta$, then the unit tangent vector $u_r$ to the surface at $(\varphi, \eta)$ which makes an angle $+ \pi/2$ with the axial vector $$u_s= \sin\theta\, R^{-1} \partial_{\varphi} + \cos \theta\, \partial_{\eta}$$ is given by 
$$u_r = -\cos\theta \, R^{-1}\partial_{\varphi} + \sin\theta\,\partial_{\eta}.$$
 Hence the angle $\vartheta$ between $F_T$ and the 
perpendicular $u_r$ to the dumbbell is given by 
$$\cos\vartheta = \cos\alpha\sin\varphi\cos\theta - \sin\alpha \sin\theta.$$
Consider the manifold $N = (-\pi/2,\,\pi/2)\times S^1$ with a projection map 
$q: M\to N$ which sends a point $(\varphi,\eta,\theta)$ to the point $(\varphi,\theta)$. As $\vartheta$ is a function of $\varphi$ and $\theta$ (but independent of $\eta$), it descends to a function on $N$, which we again denote by $\vartheta$. 

Let $S \subset N$ be the curve defined by the equation $\vartheta =\pi/2$ 
	(see Fig.\ref{fig:S_curve_torque}) \label{Def:S}. In physical terms, 
a dumbbell lies along a flow line of $\partial/\partial z$ 
(these are the red curves in Fig.\ref{fig:coordinates}) if and only if the point defined by it in $N$ lies on $S$. Such a dumbbell will not roll even if $\mu^{stat}_r=0$.  There is a certain subset $A \subset N$, which is defined by the following property. A stationary dumbbell remains stationary if and only if it is represented by a point in $A$. In equational terms, $(\varphi, \theta) \in A$  if and only if we have 

\begin{enumerate}[resume]
	\item  $|\vartheta| \le  \pi/2 - \epsilon^{stat}$ and $|F_T/F_N| < \mu^{stat}_r/\cos\vartheta$, or
	\item $|\vartheta| \ge \pi/2 - \epsilon^{stat}$  and $|F_T/F_N| < \mu^{stat}_s$. 
\end{enumerate}

The region $A$ has a subset $B$ which has the property that a stationary dumbbell whose corresponding point lies in $B$ remains stationary even when given a small nudge. It is defined in equational terms by replacing in the definition of $A$ the static friction coefficients $\mu^{stat}_r$ and $\mu^{stat}_s$ (and the resulting quantity $\sin \epsilon^{stat}$) by their dynamic analogs $\mu^{dyn}_r$ and $\mu^{dyn}_s$ (and the resulting quantity $\sin \epsilon^{dyn}$). As the coefficients of dynamic friction are less than the corresponding coefficients of static friction, $B$ is strictly contained in $A$.  These regions are depicted in Fig.\ref{fig:s-curve_schematic}.

The complement of $A$ in $N$ is defined by the property that if a stationary dumbbell is placed on the cylinder such that the corresponding point $P$ lies in $N -A$, then the dumbbell  begins to move. The nature of this motion can be quite different depending on where  the point $P$ lies within the region  $N-A$.

\subsubsection*{The torque on a dumbbell}
As we have seen, the subset $M_{min} \subset M$, defined by $\varphi =0$ and $\theta =0$,
is the locus where the potential energy of a single dumbbell in a horizontal cylinder 
attains its absolute minimum. As such, it is an attractive fixed point set. 
Energy minimization thus affects $\theta$, taking it towards $\theta=0$, which means there is
torque which reduces $|\theta|$ to $0$. This torque originates from the finite length of the dumbbells,
because of which the lower ball of the dumbbell (where $|\varphi|$ is smaller) experiences a 
higher reactive force from the cylinder as the slope of the cylinder goes to $0$ with $|\varphi|$. 
Hence the two balls experience different forces, including different tangential and normal forces, 
and so also a different frictional resistance to motion. 
The torque is zero also when  $|\theta| =  \pi/2$, which however is a repulsive fixed set. 

When a cylinder is tilted,   
the values of the forces change, and so the above torque changes.
However, when $\alpha \le 10^{\circ}$,
for small values of $|\theta|$ the effect of the tilting on the torque is small, not changing the qualitative conclusion that 
$|\theta|$ will tend to $0$ for an isolated dumbbell. If $|\theta|$ is large, the torque tends to increase $|\theta|$ towards $\pi/2$ (see Fig.\ref{fig:S_curve_torque}). In any case,
a dumbbell with $|\theta|$ large rolls down rapidly to the bottom of the tilted cylinder.

Observation shows that the tilting of a rotating cylinder (with $\alpha = 7^{\circ}$)
does not much affect the values of the stable headings dyads
as compared with the values for horizontal case
($\alpha =0$).

\subsubsection*{Turning points of zigzag paths}
Note that the regions $A$ and $B$ are invariant under the involution
$N\to N : (\varphi, \theta) \mapsto (-\varphi, -\theta)$. In particular, the respective boundaries $\partial A$ and $\partial B$ also have this involutive symmetry. However, this
symmetry gets broken by the rotational motion of the cylinder, which singles out a special subset $\partial^+ A$ (the `eastern boundary') of $\partial A$.  The subset $\partial^+ A$ consists of all
$(\varphi, \theta) \in A$ such that for any sufficiently small positive real number $\epsilon$,
the displaced point $(\varphi + \epsilon, \theta)$ does not lie in  $A$.
In words, a point of $A$ lies in $\partial^+ A$ if and only if the rotation of the cylinder almost immediately carries it outside $A$. These subsets are depicted in Fig.\ref{fig:s-curve_schematic}.
For a rotating cylinder, the phrase `stationary dumbbell' lying on the cylinder 
will mean a `relatively stationary dumbbell' lying on the cylinder, that is, one
for which the instantaneous relative velocity is zero. 
When the rotation of the cylinder carries a stationary dumbbell with a small value of $|\varphi|$ 
and $|\theta|$ to a new point of $M$ such that its
image in $N$ crosses $\partial A$, the dumbbell begins to roll down, till its corresponding point in $N$ 
enters the region $B$, when it comes to a stop after losing its momentum. This process iterates itself, leading
to the zigzag paths in $\Omega$.  The experientially obtained turning points of such zigzags are plotted in the space $ N $ in Fig.\ref{fig:s-curves}. These data lie on the expected regions $ \partial^{+} A $ and $ B $.

%
%

%
%


\subsection{Curvilinear polyhedra}\label{appn_polyhedron}

\begin{figure}[t]
	\centering
	\includegraphics[width=.75\linewidth]{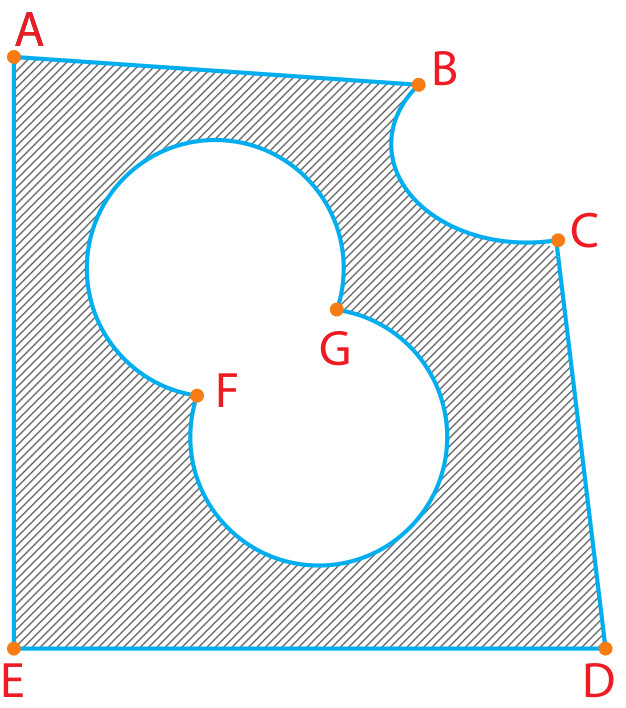}
	\caption{ The figure shows a   locally convex curvilinear polyhedron  in $\R^2$, in which  $D_0=\{A,B,C,D,E,F,G\}$, $D_1$ consists of  seven blue segments and arcs, while $D_2$ is the 
		gray shaded open region. This example has been chosen to be   locally convex  without being globally convex or even locally convex in the Euclidean sense. Here, $ D_0 $ is the set of 
		smallest dimensional corners. } 
	\label{fig:tutorial_polyhedral}
\end{figure}

Recall that a convex linear polyhedron $D$ in an affine space $\R^n$ is an intersection of finitely many
linear half-spaces $D = H_1\cap \ldots\cap H_r$. Here, each $H_i$ is defined
by a linear inequality $\sum_{j=1}^n\,a_{i,j}x_j - b_i \ge 0$ where $x_j$ are the cartesian coordinates on $\R^n$
and the $a$'s and $b$'s are constant with not all $a$'s zero. 
More generally, a linear polyhedron (not necessarily convex) $D$ in the 
affine space $\R^n$ is a finite union of convex linear polyhedra as defined above. 

We are interested in a geometric object $D$ which lives in a manifold $X$ of dimension $n$,
which may be locally regarded as a curvilinear version of the above.
A {\it locally curvilinear polyhedron} $D$ in  $X$ (a `polyhedron' for short)
is any closed subset $D$ of $X$ which satisfies the following condition: the manifold $X$ should admit
an open cover by subsets $U_{\lambda}$ together with diffeomorphisms
$\phi_{\lambda} : U_{\lambda} \to U'_{\lambda}$ where $U'_{\lambda}$ is an open set in $\R^n$,  
and for each $\lambda $ a polyhedron $D'_{\lambda} \subset \R^n$ such that $\phi_{\lambda} (U_{\lambda}\cap D) = U'_{\lambda}\cap D'_{\lambda}$.
If we can so choose the data that each $D'_{\lambda}$ is a convex linear polyhedron 
in $\R^n$, then we will say that $D$ is a {\it locally convex curvilinear polyhedron} in the manifold $X$.

It should be noted that the manifold $X$ is not assumed to  be riemannian, and
the word `convex' comes from the local diffeomeorphism with $U'_{\lambda}\cap D'_{\lambda}$, and not from any notion of convexity based
on geodesics.
		
Any  polyhedron $D'\subset \R^n$ is a disjoint union of subsets 
$$D' = D'_0 \cup \ldots \cup D'_n$$
where $D'_0$ is the set of all its vertices, $D'_1$ is the union of its 
edges, etc. In particular, $D'_n$ is the interior of $D'$. This allows us to decompose a
polyhedron $D\subset X$ similarly as a disjoint union 
$$D = D_0 \cup \ldots \cup D_n$$
in a well-defined manner, independent of the choice of the local  diffeomerphisms $\phi_{\lambda}$.
Note that the boundary of $D_r$ is contained in the union of the lower $D_i$, in fact,
$$\partial D_r  = D_0 \cup \ldots \cup D_{r-1}$$
when $D$ is connected. 
It follows from the definition that each $D_r$, if non-empty,  is a locally closed submanifold of $X$ of dimension $r$. In particular, if $r$ is the smallest integer such that $D_r$ is not empty, 
then $D_r$ is closed in $X$. At the other end, the stratum $D_n$ is open  in $X$, being the interior of $D$. Of course, $D_n$ can be empty.
The {\it dimension} of $D$ is equal to the largest $r$ for which $D_r$ is nonempty. 
The figure Fig.\ref{fig:tutorial_polyhedral} shows an example of a 
$2$-dimensional  curvilinear polyhedron in the $2$-dimensional ambient manifold $\R^2$.

\subsection{Energy minimization over a polyhedron}\label{appn_enrgy}
The main example of a locally  convex curvilinear polyhedron for us is the 
configuration space $D\subset M\times M$ for a pair of dumbbells in a cylinder that we have physically
described in section 5 above. We assume that the cylinder is kept horizontal. 
Let $E$ denote the resulting gravitational potential function on $M\times M$ which is the sum of the potential energies of the individual dumbbells.
We now consider the problem of finding the loci of local minima for 
the restriction $E|_D$ of $E$ to $D$. For simplicity, we will assume that the rod of the dumbbell
is weightless. If the rods have a weight, the value of $\theta_s$ will go up. For our experimental parameters, the effect of the weight of the rods is negligible: see Fig.\ref{fig:length_sat}. 
As a local minimal point for $E|_D$ may lie on the boundary of $D$, where $D$ is not a manifold locally, the usual calculus method of finding stationary points via vanishing of 
the first derivative needs to be replaced by more general statements. 
If a polyhedron $D \subset X$ is locally a quadrant around a minimal point, that is, 
if $D$ can be locally defined by inequalities $x_i\ge 0$ for $1\le i\le p$ for
some $p\le n$ where $x_i$ are local coordinates centered at a minimal point, then we can apply 
the following minimization lemma. It turns out that this is enough for our purpose, because
$D\subset M\times M$ is indeed locally a quadrant around the corner $\Delta$ as we see later.

%

\medskip

\centerline{\it Minimization in the corner of a quadrant}

\medskip


\noindent{\bf Lemma} 
{\it	Let $n = p+q+r$. 
	Let $C\subset \R^n$ be any subset such that the origin $0$ lies in $C$ and such that for any
	$(x_1,\ldots,x_n)\in C$ the condition $x_i \ge 0$  is satisfied
	for all $1\le i\le p $. Let $f$ be a smooth function
	in a neighbourhood of $0$ in $\R^n$, which satisfies the following properties.\\
	(1) $\partial f/\partial x_i|_0 > 0$ for $i = 1,\ldots,p$.\\
	(2) $\partial f/\partial x_j|_0 = 0$ for $j = p+1,\ldots,p+q$.\\
	(3) The $q\times q$ Hessian matrix
	$[\partial^2 f/\partial x_j\partial x_k|_0]$, where $j,k \in [p+1, p+q]$, is positive definite.\\
	(4) $f$ is independent of the remaining $r$ variables $x_{p+q+1},\ldots, x_n$.\\
	Then the point $0\in C$ is a point of local minimum
	for the restriction $f|_C : C\to \R$ of $f$ to $C$.
	Moreover, 
		the set $C\cap Z$, where $Z$ is defined by the equations
		$x_{p+q+1} = \ldots =  x_n =0$, is a locus of local minimum for $f|_C$
		in a neighbourhood of $0$.
}

\begin{figure}[t]
	\centering
	\includegraphics[width=1\linewidth]{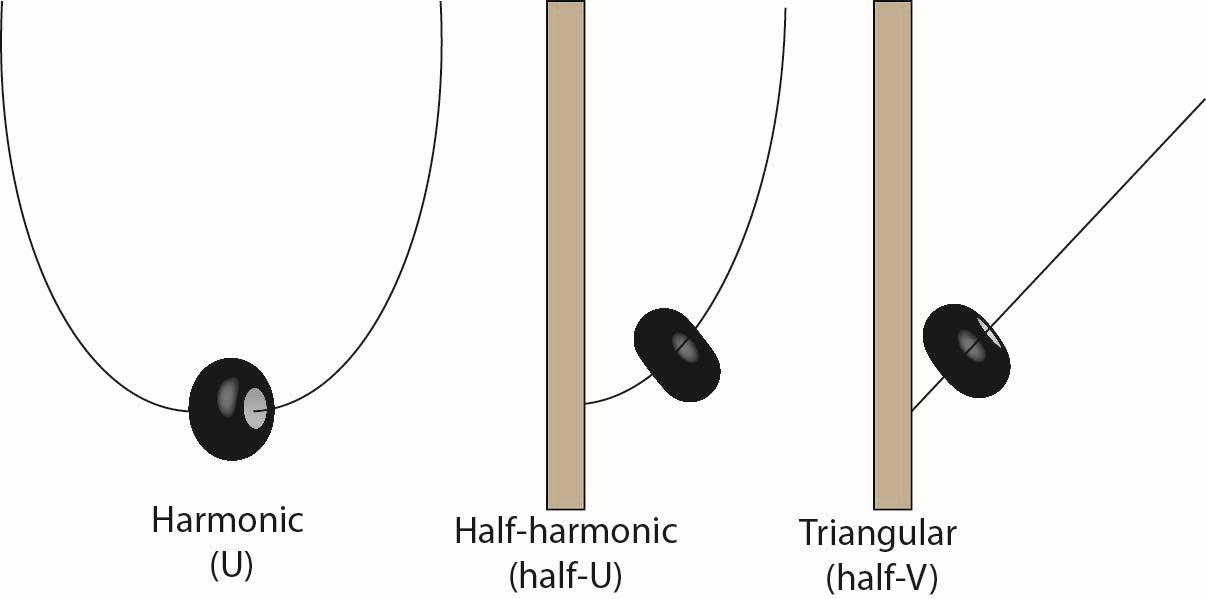}
	\caption{Varieties of entrapment}
	\label{fig:entrapment}
\end{figure}
We will apply the above by taking $C$ to be the portion of a polyhedron $D$ in a neighbourhood of the
origin $0$, such that each $x_i$ is non-negative on $D$ for $i= 1,\ldots,p$.
It is significant that the above conditions for minimization of $f|_D$
allow some of the 
first derivatives (namely, $\partial f/\partial x_i$ for $1\le i\le p$) to be positive. This is in contrast to the
minimization condition for $f$ on a manifold $X$, where {\it all} first derivatives need to be zero.
This non-vanishing of $\partial f/\partial x_i$ means that the force field $-{\rm grad}(f)$ is 
{\it non-zero} at a point $P \in D\cap Z$, and so reinforces the entrapment of the system in the corner.
In contrast, at the usual kind of minimization on manifolds, the first derivative is zero and second
derivative is positive, and so $-{grad}(f)$ is zero at the point itself, so the 
entrapment is much weaker. 

These considerations lead to a qualitative description of any higher dimensional entrapment as a combination of three  kinds of  basic one-dimensional entrapments, each of which is stronger than its predecessor, which are as follows: see Fig.\ref{fig:entrapment} for a pictorial description.

\noindent{(1)} {\it Harmonic localization ( `U' type entrapment).} This results when the configuration space is one dimensional, and contains an open interval 
$(-a,a)$ around $x=0$. Suppose the leading term of the potential energy is $x^2$. Then the system admits
a bound state around $x=0$. 

\noindent{(2)} {\it Half-harmonic (`half-U') entrapment.} This refers to a one dimensional polyhedron which is 
locally a neighbourhood of $0$ in the positive half line $x\ge 0$. Again let the leading term of the potential energy be 
$x^2$. Let the boundary at $x=0$ be an energy absorbing boundary. Due to repeated losses of energy at $x=0$ the system gets entrapped near the boundary.  

\noindent{(3)} {\it  Triangular (`half-V') entrapment.} This again refers to a one dimensional polyhedron which is 
locally a neighbourhood of $0$ in the positive half line $x\ge 0$. But this time, let the leading term of the potential energy be 
$x$. Once again, let the boundary at $x=0$ be an energy absorbing boundary. Due to repeated losses of energy at $x=0$ the system gets entrapped near the boundary.  This entrapment is further strengthened by the fact that the force (which equals
minus the gradient of the energy) is non-zero at $x=0$. Consequently, the system is rapidly driven into the `corner' point
$x=0$, and is held there with a non-zero force.

In all the above three cases, presence of friction can further enhance the entrapment.

We now apply the above lemma to our case of dumbbell pairs, to establish the following:
\label{prop:local minima} The sets $\delta_L$ and $\delta_R$ in $M\times M$ are sets of local minima for
the	energy function $E|_D$. 

To put this in the notation of the lemma, 
consider the following local coordinates $x_1,\ldots,x_6$ on $M\times M$ in a neighbourhood of a point of $\delta_L$ or
$\delta_R$. Let
\begin{eqnarray*}
x_1 &=& d(A_2,B_1) - r,\\
x_2 &=& d(\alpha, B_1) -r -t,\\ 
x_3 &=& d(A_2,\beta) - r -t ,\\
x_4 &=& \varphi ,\\
x_5 &=& \theta \pm \arctan(r/\ell), \mbox{ and}\\
x_6 &=& \eta + const. 
\end{eqnarray*}
where $\varphi$, $\theta$ and $\eta$ the average coordinates for the pair.
In the equation for $x_5$, the minus sign is for $\delta_L$ and the plus sign is for $\delta_R$.  
Then a straightforward calculation shows that the  conditions 
in the statement of the above lemma are satisfied at $(x_1,\ldots,x_6) = (0,\ldots,0)$
with $n =6$, $p = 1$, $q =4$ and $r=1$.
As explained earlier, the subset $D \subset M\times M$ is indeed given in a neighbourhood
of a point of $\delta_L$ or $\delta_R$ by the inequalities $x_i \ge 0$ for $i= 1,2,3$. 
Hence the lemma applies to give our desired conclusion. In fact, we now see that 
the entrapment in the $x_1$ direction is of the half-V type, in the $x_2$ and $x_3$ directions
it is of half-U type, in the $x_4$ and $x_5$ direction it is of U-type, while in the $x_6$ direction,
which is along $\delta$, there is no entrapment as the system is isoenergetic.

\bigskip

\noindent{\bf Acknowledgement} The authors thank Rajaram Nityananda for his careful reading of an earlier version, 
and making suggestions to improve the exposition. 
\newpage

\end{document}